\documentclass[useAMS,usenatbib]{mn2e}

\usepackage{graphicx}
\usepackage{txfonts}
\usepackage{upgreek}
\usepackage{epstopdf}
\usepackage[normalem]{ulem}
\usepackage{natbib}
\usepackage{color}
\usepackage{xcolor, cancel}
 
\title[Evolution and characteristics of forced shear flows]{Evolution and characteristics of forced shear flows in polytropic atmospheres:  Large and small  P\'eclet number regimes}
\author[V. Witzke, L. J. Silvers and B. Favier]{V. Witzke$^{1, 2}$\thanks{E-mail:Veronika.Witzke.1@city.ac.uk (VW); Lara.Silvers.1@city.ac.uk (LJS); Favier@irphe.univ-mrs.fr (BF)}, L. J. Silvers$^{1}$ and B. Favier$^{3}$ \\
$^{1}$Department of Mathematics, City, University of London,
              Northampton Square, London, EC1V 0HB, UK\\
$^{2}$Max Planck Institute for Solar System Research, Justus-von-Liebig-Weg 3, 37077 G\"ottingen, Germany \\
$^{3}$Aix-Marseille Universit\'{e}, CNRS, Ecole Centrale Marseille, IRPHE UMR 7342, 49 rue F. Joliot-Curie, 13013 Marseille, France}
\begin{document}
\newcommand\redsout{\bgroup\markoverwith{\textcolor{red}{\rule[0.5ex]{2pt}{0.4pt}}}\ULon}
\maketitle
\label{firstpage}
\begin{abstract}
Complex mixing and magnetic field generation occurs within stellar interiors particularly where there is a strong shear flow. To obtain a comprehensive understanding of these processes,  it is necessary to study the complex dynamics of shear regions. Due to current observational limitations, it is necessary to investigate the inevitable small-scale dynamics via numerical calculations. Here, we examine direct numerical calculations of a local model of unstable shear flows in a compressible polytropic fluid primarily in a two-dimensional domain, where we focus on determining how key parameters affect the global properties and characteristics of the resulting saturated turbulent phase. We consider the effect of varying both the viscosity and the thermal diffusivity on the non-linear evolution. Moreover, our main focus is to understand the global properties of the saturated phase, in particular estimating for the first time the spread of the shear region from an initially hyperbolic tangent velocity profile. We find that the vertical extent of the mixing region in the saturated regime is generally determined by the initial Richardson number of the system. Further, the characteristic quantities of the turbulence, i.e.~typical length-scale and the root-mean-square velocity are found to depend on  both the Richardson number, and the thermal diffusivity. Finally, we present our findings of our investigation into saturated flows of a `secular' shear instability in the low P\'eclet number regime  with large Richardson numbers.
\end{abstract}
\section{Introduction}
Developing a complete model of stellar dynamics requires a comprehensive understanding of the microphysical and macrophysical processes present in all stellar regions. Current stellar evolution models are challenged by some discrepancies between theory and observations, which can only be resolved by introducing additional mixing \cite[see][and references therein]{doi:10.1146/annurev.astro.35.1.557}. In stars, a possible source for such mixing processes is shear-induced turbulence (\citealp{Zahn_1974, 1977A&A....56..211S, 1978ApJ...220..279E}). In addition, complex gas dynamics in stellar interiors is not only important for mixing processes but also plays a crucial role in magnetic field generation \cite*[see][]{annurev.010908.165215, 5188151520100501}. Therefore, ongoing research focuses on understanding possible hydrodynamical and magnetohydrodynamical instabilities leading to turbulence in various stellar regions \citep*{2012MNRAS.425.2267R, 2015AandAWitzke, 2016ApJ...821...49G}.\\   
%
%
Main-sequence stars have common dynamical elements, such as large-scale shear flows resulting from differential rotation. However, these shear flows can occur in different regions with different characteristics i.e.\ different transport coefficients, thermal stratification, etc. One important body is the Sun, where considerable effort has recently been directed towards extending our understanding of the tachocline and its role in the solar dynamo \cite[see][and references therein]{2008RSPTA.366.4453S}. In the Sun there is also the near-surface shear layer that is located at the upper boundary of the convection zone. The near-surface shear layer is believed to be important for the magnetic field generation \citep{2005ApJ...625..539B,Cameron_sch_2017}, but it has significantly different transport coefficients (\citealp{Thompson1300, 2011ApJ...743...79M}; \citealp*{2014A&A...570L..12B}). Thus non-dimensional numbers such as the Reynolds number or the P\'eclet number can vary in astrophysical objects. The P\'eclet number, which is the ratio of advection to temperature diffusion, plays an important role in the linear and non-linear stability of shear flows. In particular, a low P\'eclet  number, i.e. less than unity, affects the stability threshold of shear flows \citep{Zahn_1974, 1999AA...349.1027L, Garaud10.10631.4928164}, as it destabilises the system and facilitates an instability. Such a low P\'eclet number regime can be reached in stellar regions where there is high thermal diffusivity, which can occur deep in stellar interiors or in envelopes of massive stars \citep{2016ApJ...821...49G}. \\
%
In the classical stability analysis of a vertical shear flow in an ideal fluid, the Richardson number, which corresponds to the ratio of the  Brunt-V{\"a}is{\"a}l{\"a} frequency and the turnover rate of the shear, gives the measure for stability.  A low Richardson number, less than 1/4, is required for the system to become unstable. However, in diffusive systems, where the P\'eclet number is low, a possible `secular' shear instability can be present. This type of shear flow instability develops even for high Richardson numbers but only if the thermal diffusivity is large enough to weaken the stable stratification \citep*{Zahn_1974, 1999AA...349.1027L, Garaud10.10631.4928164, 2015AandAWitzke}.  So far investigations of low P\'eclet number shear flows have made use of the Boussinesq approximation \citep{2016A&A...592A..59P}, which does not permit the study of a system that is larger than a pressure scale height as is present in stellar interiors. \\
%
Numerical investigations of the dynamics in stellar interiors are challenging because the viscosity as well as the Prandtl number present in such regions are very low, which leads to high Reynolds numbers and low P\'eclet numbers on small scales. The considerable computational cost of such numerical calculations is not accessible with current resources.  Using  considerably larger values of the Prandtl number, or the viscosity, in order to obtain a tractable system is a way forward but this will necessarily lead to some differences in the dynamics that would be found in  real astrophysical shear flows. Moreover, the Richardson number in the tachocline is approximated to be  $\mathcal{O}(10)$. While this approximation is obtained from spatially and time-averaged measurements, turbulent motions can be present on smaller length-scales and time-scales.\\
Recent investigations have focused on the conditions under which low P\'eclet number flows become linearly or non-linearly unstable \citep[see][]{2013A&A...551L...3P, Garaud10.10631.4928164, 2016ApJ...821...49G} and mixing can occur. However, the relevant parameters that determine whether thermal diffusivity has significant impact on the non-linear dynamics of the system are dictated by the typical length-scales and velocities in the turbulent regime and so they are not known \textit{a priori}, but can only be determined from fully non-linear hydrodynamical calculations. In stellar interiors these parameters evolve due to complex mechanisms, which are not entirely understood. Moreover, previous numerical studies used a periodic domain in the vertical direction \citep{2016ApJ...821...49G} or a linear shear profile \citep{2014A&A...566A.110P}. Both approaches correspond to modelling a very localised part of a larger shearing region.
While they have extended our understanding, it remains unclear if the findings of these approaches persist when investigating shear transition regions in their entirety.\\
%
In this paper we examine a local region of a fully compressible, stratified fluid to improve our understanding of shear regions in stellar interiors.
Using a hyperbolic tangent velocity profile permits us to model a larger region of differential rotation, where a sharp shear flow is localised at the middle.  Such a shear flow allows for different dynamics to occur compared to the aforementioned studies.  Moreover, our setup is not periodic in the vertical direction and highly stratified, which differs significantly from a standard approach where all dimensions are periodic.
Our investigations address the following questions: Does the viscosity have an impact on the spread of a shear flow instability and how are the turbulent length-scales affected?  What effect does the thermal diffusivity  have on the resulting  shear region and turbulence characteristics present there? To what extent is the spread of the shear instability controlled by the Richardson number? \\
While we expect viscosity to have an effect on the smallest possible length-scales it  is not obvious if the typical length-scale will be significantly affected. 
We will primarily focus on  classical shear instabilities in a two-dimensional setup with different P\'eclet numbers in order to investigate the resulting saturated regime and what affects the extent of the shear region after saturation.\\
%
The paper will proceed  as follows. In Section \ref{sec:model} the governing equations are given together with the numerical methods used. In Section \ref{sec:results_part2} we examine how unstable shear flows saturate and evolve into quasi-static states when transport coefficients are changed. Focusing on a low Richardson number instability, an overview of the effects of varying the Reynolds number and initial P\'eclet number is given. Finally, the possibility of a `secular' instability is investigated, where large Richardson numbers are considered. Here, the non-linear regime is compared to a low P\'eclet number turbulence induced by a classical low Richardson number instability.
%
\section{Model}
\label{sec:model}
%
We consider an ideal monatomic gas with constant dynamic viscosity, $\mu$, constant thermal conductivity, $\kappa $, constant heat capacities $c_p$ at constant pressure and $c_v$ at constant volume, and with an adiabatic index $\gamma= c_p/c_v = 5/3$. In  this study we  chose  to fix the dynamic viscosity rather than the kinematic viscosity.  The domain throughout most of this paper is a  $x$-$z$ plane, where the horizontal \textit{{x}}-direction is periodic and the depth in the $z$-direction is $d$.  The depth of the domain in dimensional units is given by $\tilde{z}$, where the two boundaries  are located at $\tilde{z} =0$ and $\tilde{z}= d$. Most of our calculations were performed in this two-dimensional setup in order to reduce the computational cost.  Thus we consider a two-dimensional set of differential equations, i.e.~we use three-dimensional arrays, but we neglect the spatial variations in a possible third $y$-direction, which is perpendicular to the $x$-$z$ plane, and assume that all quantities are zero in this possible third direction. \\
The full set of dimensionless,  differential equations is:
\begin{eqnarray}
\label{eq:NSEquation01}
\frac{\partial \rho}{\partial t} & = & - \mathbf{\nabla}\mathbf{\cdot}\left(\rho \mathbf{u} \right) \, \\
\label{eq:NSEquation002}
\frac{\partial(\rho \mathbf{u})}{\partial t} & = & \sigma C_k \left( \nabla^2\mathbf{u}\, +\,\frac{1}{3}\mathbf{\nabla}(\mathbf{\nabla}\mathbf{\cdot}\mathbf{u})\right) -\mathbf{\nabla} \mathbf{\cdot} \left(\rho \mathbf{u u} \right)\, \nonumber \\
 &   & -\,\mathbf{\nabla}p\, + \, \theta(m+1) \rho\, \hat{\mathbf{z}} \, + \, \mathbf{F}\, \\
\label{eq:NSEquation02}
\frac{\partial T}{\partial t}  & = & \frac{C_k \sigma (\gamma -1)}{2\rho}|\mathbf{\tau}|^{2}\,+\,\frac{\gamma C_k}{\rho} \nabla^2 T \nonumber \\
 &  & - \mathbf{\nabla} \mathbf{\cdot}\left(T \, \mathbf{u}\right)\,-\,(\gamma -2)T \mathbf{\nabla} \mathbf{\cdot} \, \mathbf{u}
\label{eq:NSEquations}
\end{eqnarray}
where $\rho$ is the density, $\mathbf{u}$ the velocity field, $T$ the temperature, $p$ is the pressure, $\theta$ denotes the  temperature difference across the layer and $\hat{\mathbf{z}}$ is the unit vector in the $z$-direction. 
In the dimensionless equations above, all lengths are given in units of the domain depth $d$, where $z = \tilde{z}/ d$ is the non-dimensional vertical length. The temperature and density are recast in units of $T_{t}$ and $\rho_{t}$, the temperature and density at the top of the layer, and we take the sound-crossing time, which is given by $\tilde{t} = d/[ (c_p-c_v)T_{t} ]^{1/2}$, as the reference time. 
There are two additional dimensionless numbers in the set of equations above: the Prandtl number, $\sigma=\mu c_p/ \kappa $, which is the ratio of viscosity to thermal  diffusivity and the thermal diffusivity parameter $C_k= \kappa \tilde{t} /(\rho_t c_p d^2)$.  The strain rate tensor in equation (\ref{eq:NSEquation02}) has the form 
 \begin{equation}
 \tau_{ij}=\frac{\partial u_{j}}{\partial x_i} + \frac{\partial u_i}{\partial x_j} - \delta_{ij} \frac{2}{3}  \frac{\partial u_k}{\partial x_k} , 
 \end{equation} 
where its tensor norm is $|\mathbf{\tau}|^{2} = \tau_{ji} \tau^{ij}$. 
For the basic state a polytropic relation between pressure and density is taken. 
In this paper the polytropic index $m$ always  satisfies the inequality $m > 1/(\gamma -1) = 3/2$, such that the atmosphere is stably stratified. Note, that the gravitational acceleration, $g$,  is derived from the hydrostatic equilibrium. It is assumed constant throughout our domain due to the local assumption of a polytropic atmosphere, which leads to $g= \theta (m + 1)$, and the Cowling approximation \citep{1941MNRAS.101..367C}.  Our parameter choices for all calculations presented in this paper are summarised in Table \ref{table:two_dim_surv} and Table \ref{table:secular_Instability}.\\
The boundary conditions at the top and the bottom of the domain are impermeable and stress-free velocity and fixed
temperature: 
\begin{eqnarray}
\label{eq:boundary01}
{u}_z = \frac{\partial {u_x}}{\partial z}  = 0  \qquad \textrm{at} \qquad  z = 0   \quad \textrm{and} \quad  z = 1, \\
T=1 \quad \textrm{at} \quad z=0 \quad \textrm{and} \quad  T=1+\theta  \quad \textrm{at} \quad z=1.
\end{eqnarray}
The dimensionless initial temperature and density profiles are of the form:
\begin{equation}
T(z)= \left(1 + \theta z \right)
\end{equation}
\begin{equation}
\rho (z) =  \left( 1 + \theta z \right)^{m}.
\end{equation}
This basic state corresponds to an equilibrium state if the fluid is at rest. However, we assume that an external force, denoted $\mathbf{F}$ in equation (\ref{eq:NSEquation002}), sustains the following initial background velocity profile 
\begin{equation}
\label{eq:target_shear}
\mathbf{U_0} = ( u_0(z), 0, 0 )^T = \frac{U_0}{2} \tanh\left(\frac{2}{L_u}{(z-0.5)}\right) \hat{\mathbf{e}}_x
 \end{equation}
where  $U_0$ is the shear amplitude and $L_u$ is the width of the shear profile. A hyperbolic tangent shear profile was chosen to minimize the boundary effects. Although the relevant parameter values for the investigations were chosen to keep the spread of the instability confined in the middle domain, it is impossible to avoid completely  any boundary effects. Furthermore, the boundary conditions introduced in equation~(\ref{eq:boundary01}) restrict the shear profile to values of $L_u$ that will result in a low enough value of the \textit{z}-derivative of $u_0$ at the boundaries.  
A visualisation of the general form of shear, density and temperature profiles used in this paper can be found in \citet*{2015AandAWitzke}.\\
The force term, $\mathbf{F}$, in equation (\ref{eq:NSEquation002}) aims to model external forces 
resulting from large-scale global effects (such as Reynolds stresses associated with thermal convection in global-scale calculations for example) that are not included in our local approach. In order to balance the viscous dissipation associated with the initial shear flow profile given by equation (\ref{eq:target_shear}), the force  
\begin{equation}
\label{eq:viscous_force}
\mathbf{F} = - \sigma C_k \nabla^{2} \mathbf{U}_0
\end{equation}
was included in equation (\ref{eq:NSEquation002}). For the set of equations (\ref{eq:NSEquation01})-(\ref{eq:NSEquations}) with the viscous forcing given by equation (\ref{eq:viscous_force}) the basic state described above with the initial velocity profile is only an equilibrium state if viscous heating, the first term on the right-hand side of equation (\ref{eq:NSEquations}), is neglected. For all the cases that we consider, the time-scale of the shear instability is at least two orders of magnitude smaller than the viscous heating time-scale of the system. Thus there is a time-scale separation between the growth rate of the instability and the viscous evolution of the shear flow, so that assuming we have an equilibrium is reasonable. \\
This method has been broadly applied to model forced shear flows and the dynamics of the solar tachocline \citep[e.g.][]{0004-637X-586-1-663, 1538-4357-702-1-L14}. Note that this method only balances  the viscous diffusion of momentum associated with the target profile and does not depend on the actual non-linear solution.  It is always the case that any kind of  forcing method has an effect on the saturated regime. Moreover, most forcing methods reach a state where the injected energy and the dissipated energy are in balance. We chose a method that ensures a balance and provides a local forcing that has minimal effect on the changes of the background profile induced by the instability. Thus it is suitable to study the characteristics of turbulent motion that are triggered by an instability in a viscous fluid. For a more detailed discussion of the appropriate forcing method see \citet*{Witzke21112016}. \\
%
%
Our calculations were initialised by adding a small random temperature perturbation to the equilibrium state including the additional shear flow in equation (\ref{eq:target_shear}). In order to evolve the system in time, equations (\ref{eq:NSEquation01})-(\ref{eq:NSEquations}) were solved using a hybrid finite-difference/pseudo-spectral code  \cite*[see][]{FLM:340261, 2009MNRAS.400..337S, FLM:8458223, FLM:8885996}. 
In addition to conducting fully non-linear direct numerical calculations, we also considered the linear stability analysis of this system, as detailed in \cite*{2015AandAWitzke}. For this the eigenvalue-problem was numerically solved on a one-dimensional grid in the $z$-direction that is discretised uniformly, and this method is adapted from \citet{2012MNRAS.426.3349F}.\\
%
%
For the characterisation of the initial state it is convenient to introduce several dimensionless numbers. 
Applying our nondimensionalisation on the Brunt-V{\"a}is{\"a}l{\"a} frequency, as derived in \citet[p.~33]{andrews2000introduction}, it becomes  
\begin{equation}
\label{eq:N_frequency}
N^2(z) =  \frac{\theta\left(m+1\right)}{{T_{pot}}} \frac{\partial {T_{pot}}}{\partial z}, \,
\end{equation}
where ${T}_{pot} = T P^{1/\gamma-1} $ is the potential temperature. Then, the minimum value of the  Richardson number, $Ri$, across the layer is defined as
\begin{eqnarray}
Ri_{min} & = & \min_{0 \le z \le 1} \left(N(z)^2 \left/ \left( \frac{\partial u_0(z)}{\partial z} \right)^2 \right.  \right) \nonumber \\
 & = & \min_{0 \le z \le 1}  \left( \frac{ \theta^2  L_u^2(m +1) \left(\frac{m+1}{\gamma} - m\right)}{\left(1+\theta z \right) \left(U_0-4 u_0(z)^2/U_0 \right)^2 } \right),
\label{eq:Richardsonnumber_def}
\end{eqnarray} 
where the derivative of the background velocity profile, defined in equation (\ref{eq:target_shear}), with respect to $z$ corresponds to a local turnover rate of the shear. In most cases the minimum $Ri$ value is at $z=0.5$, but for some parameter choices where there is a  large temperature gradient, $\theta$, and a broad shear width, the minimum is shifted towards greater $z$. For selected cases the $Ri(z)$ was plotted further below in Fig.~\ref{fig:figure_3.3RiPe} and Fig.~\ref{fig:figure_diffRi}.\\
The $1/4$ criterion is a necessary, but not sufficient, requirement for instability in an incompressible fluid. Thus, in order to verify that the cases we considered are unstable, the linear stability problem was solved. We considered unstable shear flows with a Richardson number, $Ri$, less than 1/4 at a point in the domain. However, for some investigations, systems with a minimum Richardson numbers greater than $1/4$ were considered in order to study the `secular' instability. \\
Furthermore, to characterise the system we used the initial P\'eclet number at the top of the domain, i.e. $z=0$, which  we define as
\begin{equation}
\label{eq:Peclet_number_in}
Pe = \frac{ U_0 L_u}{C_k}  \rho(0) \, ,
\end{equation}
where $U_0$ and $L_u$  are as defined in equation~(\ref{eq:target_shear}). The  P\'eclet number is also useful for the so-called `secular' instabilities. In Subsection \ref{Sec:Secular_Instability} we considered shear instabilities induced by the destabilizing effect of thermal diffusion for which larger values of $Ri$ can be used \citep{Dudis_1974, Zahn_1974, 1999AA...349.1027L}.  Finally, the initial  Reynolds number, defined at the top of the domain, is 
\begin{equation}
Re =  \frac{U_0 L_u}{\sigma C_k} \rho(0).
\end{equation}
Our choice of fundamental units is useful when a fully compressible, polytropic atmosphere is studied, and naturally differs from previous Boussinesq studies \citep[see for example][]{Jones1977,  1999AA...349.1027L, doi:10.1146/annurev.fluid.35.101101.161144}. In the following we  discuss the link between the free parameters that appear in the equations (\ref{eq:NSEquation01})-(\ref{eq:NSEquations}) and  the Reynolds number, the Richardson number and the P\'eclet number, which are typically used when incompressible shear flows are studied.  Varying the Prandtl number, $\sigma$, corresponds to a change in the  Reynolds number only. The P\'eclet number can be associated with the thermal diffusivity parameter, $C_k$. However, in order to vary only the P\'eclet number,  it is necessary to keep the viscosity fixed, i.e. $\sigma$ needs to be adjusted accordingly. Finally, the Richardson number can be varied independently from $Re$ and $Pe$ by changing the temperature gradient, $\theta$, or the polytropic index, $m$. To characterise the initial setup the initial $Pe$ number, $Re$ number and minimum $Ri$ number were calculated, where the $Pe$ and $Re$ numbers were obtained at the top but using the typical length-scale from the shear profile. Since the domain is strongly stratified, and the initial configuration is a polytropic state, the pressure scale-height changes across the domain. The maximum and minimum pressure scale-heights at the bottom and top layer respectively were calculated and a rounded value is given for all the considered cases  in Table \ref{table:two_dim_surv} and \ref{table:secular_Instability}.
\section{Results}
\label{sec:results_part2}
\begin{table*}
\setlength\tabcolsep{5pt}
\centering
\caption{A comparison of typical length-scales, turbulent Reynolds numbers, $Re_t$, and effective shear width, $L_{eff}$, during the saturated phase.  The error for the effective shear width  only accounts the fitting error.  Additionally, we provide a shear width, $L_{cut}$, obtained by a cut-off method at 95\%. For all cases the polytropic index is $m=1.6$, the shear amplitude $U_0 = 0.19$, and the initial shear width is $L_u = 0.0333$. The initial P\'eclet number, $Pe$, Reynolds number, $Re$ and minimum Richardson number, $Ri$, are listed. The effective shear width is calculated after saturation and  $\bar{Re_t}$,  and $\bar{l_w}$  are averaged over the whole domain. For the cases A to H and I to L the pressure scale-heights are the same, where at the top  $H_{min} = 0.2$ and at the bottom  of the domain $H_{max}=0.6$. For the cases D1 to G3, the maximum pressure scale-height varies from $H_{max} = 0.4$ to $H_{max} = 1.2$ and for the minimum pressure scale-height from $H_{min} = 0.06 $ to $H_{min} = 0.8 $.  }
\begin{tabular}{c c c c c c c c c c c c c c  c }
\hline
Case &$\theta$  & $\sigma$ &$C_k$ &    $Pe$ & $Re$ & $Ri$  & $L_{eff}$ & $L_{cut}$&     $\bar{Re_t}$ &            $\bar{u_{rms}}$ &                  $ \max u_{rms} $    &                      $\bar{l_w}$               &  $\min l_w $  \\ 
\hline
 \multicolumn{6}{c}{Varying $Re$ via changing $\sigma$} & \multicolumn{6}{c}{Resolution $N_x = 512$, $N_z = 480$  } \\
\hline
A &1.9  & 1 &0.0001 &  60 & $6.3 \times 10^1 $& 0.006 & $0.61 \pm 0.04$ & 0.88 & $2.6 \times 10^2 $ &  $1.2 \times 10^{-2}$ & $2.4 \times 10^{-2}$  & 0.93 & 0.43  \\
B & 1.9&0.1  & 0.0001 & 60 & $6.3 \times 10^2 $& 0.006 & $0.53 \pm 0.05$ & 0.78 & $5.7 \times 10^2$ &   $2.5 \times 10^{-3}$ & $8.8 \times 10^{-3}$& 0.89 & 0.21\\
C &1.9 & 0.05 & 0.0001  & 60 & $1.3 \times 10^3 $ & 0.006 & $0.49 \pm 0.03 $ & 0.68 &  $7.3 \times 10^2$ & $1.6 \times 10^{-3}$& $6.2 \times 10^{-3}$ &0.91  &0.23   \\
\hline
\multicolumn{6}{c}{Varying $Pe$ number via changing $C_k$} & \multicolumn{6}{c}{Resolution $N_x = 512$, $N_z = 480$  }\\
\hline
D &1.9 &  0.000633 & 0.16 & 0.04& $6.2 \times 10^1$& 0.006 & $0.51 \pm 0.02 $ & 0.89    & $3.2 \times 10^2$ & $8.9 \times 10^{-3}$& $2.0\times 10^{-2}$ &1.04  &0.60\\
E &1.9 &0.00633  & 0.016  & 0.4 &$6.2 \times 10^1$&0.006 &  $0.43\pm 0.01 $&    0.86   & $1.2 \times 10^2$ &  $5.3 \times 10^{-3}$& $2.0 \times 10^{-2}$ &1.05  &0.33 \\
F &1.9 &  0.0633& 0.0016 & 4.0 & $6.2 \times 10^1$& 0.006 & $0.43 \pm 0.02$ &  0.71    & $6.7 \times 10^1$  & $4.3 \times 10^{-3}$& $1.9 \times 10^{-2}$ &0.91  &0.29 \\
G &1.9 & 0.948 & 0.00011 & 60 & $6.2 \times 10^1$& 0.006 & $0.65 \pm 0.07$ &   0.90   & $3.8 \times 10^1$   & $4.0 \times 10^{-3}$& $2.1 \times 10^{-2}$ &0.46  &0.18 \\
H &1.9 &9.48  & 0.000011 & 600 & $6.2 \times 10^1$& 0.006 & $0.95\pm 0.16$ &  0.83   & $2.4 \times 10^2$ & $1.0 \times 10^{-2}$& $2.4 \times 10^{-2}$ &0.79  &0.39 \\
\hline
\multicolumn{6}{c}{ Varying $Ri$ number via changing $\theta$} & \multicolumn{6}{c}{Resolution $N_x = 512$, $N_z = 480$  }\\
\hline
D1 &1.9 & 0.000633 & 0.16 & 0.04& $6.2 \times 10^1$  & 0.006 & $0.51 \pm 0.02 $& 0.89   & $3.2 \times 10^2$ &  $8.9 \times 10^{-3}$& $2.0 \times 10^{-2}$ &1.04  &0.60\\
D2 &4.2 & 0.000633 & 0.16 & 0.04& $6.2 \times 10^1$  &0.018 & $0.47 \pm 0.02$& 0.86    & $2.0 \times 10^2$  &  $4.0 \times 10^{-3}$& $1.6 \times 10^{-2}$ &1.12  &0.42\\
D3 &6.3 &  0.000633 & 0.16 & 0.04&$6.2 \times 10^1$   & 0.030 &$0.33 \pm 0.01$&  0.80   & $1.5 \times 10^2$ &  $2.7 \times 10^{-3}$& $1.3 \times 10^{-2}$ &0.85  &0.31\\
FF1 &0.85 & 0.0633  &0.0016 & 6 &$6.2 \times 10^1$   &$0.0006$ &$0.87 \pm 0.09$ & 0.89   & $1.7 \times 10^2$   & $1.1 \times 10^{-2}$& $3.2 \times 10^{-2}$ &1.23  &0.53 \\
FF2 &3.6 &  0.0633  &0.0016   & 6 &$6.2 \times 10^1$   &$0.006$ & $0.45 \pm 0.01 $ & 0.76   & $1.5 \times 10^2$  & $4.7 \times 10^{-3}$& $2.1 \times 10^{-2}$ &1.07  &0.26 \\
FF3 &6.5 &  0.0633  &0.0016  & 6 &$6.2 \times 10^1$   &$0.013$ & $0.23 \pm 0.01 $ & 0.67   & $3.1 \times 10^2$ & $5.0 \times 10^{-3}$& $1.6 \times 10^{-2}$ &0.68  &0.32 \\
G1 &0.49 & 0.948 & 0.00011 & 60 &$6.2 \times 10^1$   &0.0006 &$0.36 \pm 0.01$&  0.88    & $6.1 \times 10^2$   & $3.0 \times 10^{-2}$& $4.2 \times 10^{-2}$ &1.48  &1.26 \\
G2 & 1.2&0.948 & 0.00011 & 60 &$6.2 \times 10^1$  & 0.0030 &  $0.51 \pm 0.04$&  0.82  & $2.6 \times 10^2$  & $1.3 \times 10^{-2}$& $2.7 \times 10^{-2}$ &1.06  &0.60 \\
G3 &1.9 & 0.948 & 0.00011& 60 &$6.2 \times 10^1$   & 0.0060 & $0.65 \pm 0.07$&  0.90   & $3.8 \times 10^1$& $4.0 \times 10^{-3}$& $2.1 \times 10^{-2}$ &0.46  &0.18 \\

\hline
\multicolumn{6}{c}{Three-dimensional cases} & \multicolumn{6}{c}{Resolution $N_x = 256$, $N_y = 256$, $N_z =320 $  }\\
\hline 
I   &1.9 & 0.32  &  0.0016 &   4.0 & $1.2 \times 10^1$  &  0.006 & $0.29 \pm 0.01$ & 0.66   & $2.6 \times 10^1$  & $7.7 \times 10^{-3}$& $3.5 \times 10^{-2}$ &0.74  &0.36 \\
J   &1.9  & 0.016 & 0.032 & 0.2 & $1.2 \times 10^1$  &0.006 &  $0.48 \pm 0.01$ &  0.70  & $3.7 \times 10^1$ & $1.0 \times 10^{-2}$& $4.0 \times 10^{-2}$ &0.90  &0.41 \\
\hline
\multicolumn{6}{c}{Two-dimensional cases for comparison} & \multicolumn{6}{c}{Resolution $N_x = 256$, $N_z =320 $  }\\
\hline  
K  (2D) &1.9 & 0.32  &  0.0016  &   4.0 & $1.2 \times 10^1$  &0.006 &  $0.30 \pm 0.02$& 0.63    & $5.0 \times 10^1$ & $1.2 \times 10^{-2}$& $3.4\times 10^{-2}$ &0.97  &0.41 \\
L  (2D) & 1.9 &0.016 & 0.032  & 0.2 & $1.2 \times 10^1$  & 0.006  &$0.40 \pm  0.02$&  0.71   & $6.6 \times 10^1$ & $1.1 \times 10^{-2}$& $3.5 \times 10^{-2}$ &0.99  &0.51 \\
\hline

\end{tabular}
\label{table:two_dim_surv}
\end{table*}
When focusing on small-scale dynamics such as shear-induced turbulence, we would like to understand whether there is a relation between the transport coefficients, such as viscosity and thermal diffusivity,  and the characteristic length-scales and the velocities of the resulting turbulent saturated state. Furthermore, observations of relevant shear flows, such as the tachocline for example, only provide spatial and time averaged measurements \cite[see for example][]{1996ApJ...469L..61K, 1999ApJ...527..445C}. Therefore, understanding what controls the global properties of the resulting mean flow after saturation, can help to draw a connection between observations of shear flows in astrophysical objects and numerical calculations. For this we  investigated the effect of varying the values of the transport coefficients on non-linear dynamics.  Note, that the viscosity value is a limiting factor for the smallest scales that can be achieved numerically  in the turbulent regime. \\
We focus on varying $Re$ in Subsection \ref{sec:effect_of_sigma_ck}, where we also introduce all calculated quantities that are investigated for all studied cases. Then,  the $Pe$ number was varied in Subsection \ref{Sec:Vary_peclet_number} and we investigate how varying $Ri$ affects the saturated phase  in Subsection \ref{sec:Ri_Re_varying}. These investigations were conducted in two dimensions, but some representative three-dimensional cases were performed and compared to two-dimensional cases in Subsection \ref{sec:three_dim_long_time_study}. All cases that were considered are summarised in Table \ref{table:two_dim_surv}. Finally, in Subsection \ref{Sec:Secular_Instability}  we discuss the so-called `secular' shear instability \citep{1978ApJ...220..279E} in a two-dimensional setup. Here, we  investigated if the resulting state in the non-linear regime shows significantly different behaviour compared to that of a low P\'eclet number regime triggered by a `classical' shear flow instability. Parameters for the `secular' instability cases are summarised in Table \ref{table:secular_Instability}.
\subsection{The effect of varying the Reynolds number}
\label{sec:effect_of_sigma_ck} 
%
\begin{figure*}
  \vspace*{5pt}
\includegraphics[width=0.98\textwidth]{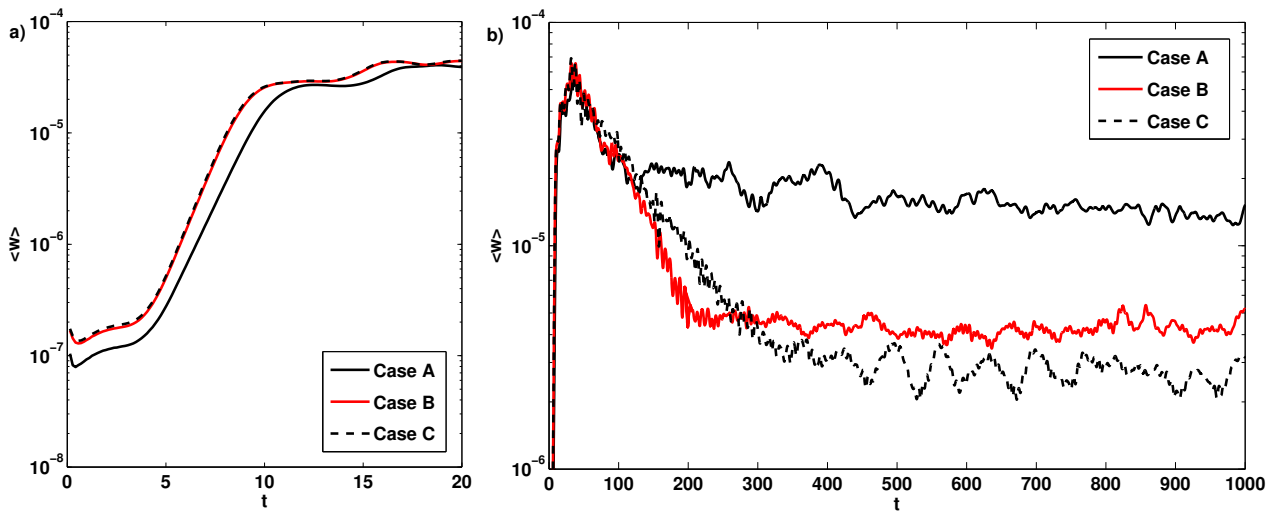}
\caption{Volume averaged vertical velocity evolution for cases A to C. In a) the early evolution, where the exponential growth is shown, whereas in b) the long-time evolution is displayed.}   
\label{fig:figure01w} 
\end{figure*}
%
In order to understand the effect of viscosity on shear induced turbulence, investigations in a  two-dimensional domain with spatial resolution $480 \times 512$ were performed. This study is summarised in Table \ref{table:two_dim_surv}. Here, a highly supercritical system, i.e.~\mbox{$Ri \ll  1/4$}, was chosen, where $Ri = 0.006$. The Mach number remains less than 0.1 throughout the domain, which was achieved by taking $U_0/2 = 0.095$, $1/L_u = 60$ and considering a temperature difference of $\theta = 1.9$. The viscosity was varied by several orders of magnitude while the thermal diffusivity was fixed in such a way that the P\'eclet number is greater than unity (cases A to C in Table \ref{table:two_dim_surv}). \\
In general, before the system reaches a quasi-static state the shear flow instability grows exponentially and evolves throughout the saturation phase. After the system  saturates, it enters a regime where statistical quantities fluctuate around a mean value. These regimes can be identified from the time evolution of the volume averaged vertical root-mean-square (rms) velocity in two dimensions
\begin{equation}
\label{eq:volume_av_w}
\left\langle w \right\rangle = \frac{1}{N_x N_z} \sum_{i=1}^{N_x} \sum_{j=1}^{N_z}  \sqrt{w(i,j)^2}.
\end{equation}
Since the overall evolution of the volume averaged vertical velocity is similar in all unstable systems, the time evolution of $\left\langle w \right\rangle$ for cases A to C is shown in Figure \ref{fig:figure01w}. We chose these three cases because they have significantly different viscosities and illustrate the effect of viscosity on the saturation. In Fig.~\ref{fig:figure01w} (a) the exponential growth of the instability is displayed, which is almost identical for the three cases because only the viscosity was changed. During the exponential growth phase, the instability growth rate does not vary for the cases in Fig.~\ref{fig:figure01w}. After approximately nine sound crossing times the system starts to saturate. This phase persists longer for lower viscosities, as can be seen in Fig.~\ref{fig:figure01w} (b).
Finally, when the volume averaged vertical velocity fluctuates around a mean value the system reaches a statistically steady state. Case A enters the statistically steady state after 150 sound crossing times, but for case C the statistically steady state is approximately after 450 sound crossing times. All statistics presented in this paper were time-averaged over a sufficiently long time interval  during the statistical steady state. In order to ensure that the system is evolved for a sufficiently long time we considered the largest diffusive time-scale $t_{{\sigma C_k},0}= 1/(\sigma C_k)$, where 1 is the non-dimensionalised length of the domain. Our calculations were evolved for at least a significant fraction of this time-scale, $> 0.2$ times the largest diffusive time-scale, which is long enough to give meaningful statistics. Note, the statistically steady state in this investigation is, to some extent, affected by the forcing method we used. The forcing method we chose here reaches a state where the work done by the forcing on the system and the viscous dissipation rate of momentum in the system are in balance \citep*{Witzke21112016}.  \\
%
We begin with a qualitative observation of the flow, by considering the vorticity component perpendicular to the \textit{x-z}-plane just after the exponential growth phase when the billows start to overturn, and during the statistical steady state. The dynamics alter significantly with decreasing viscosity. 
\begin{figure*}
\centering
\includegraphics[width=0.48\textwidth]{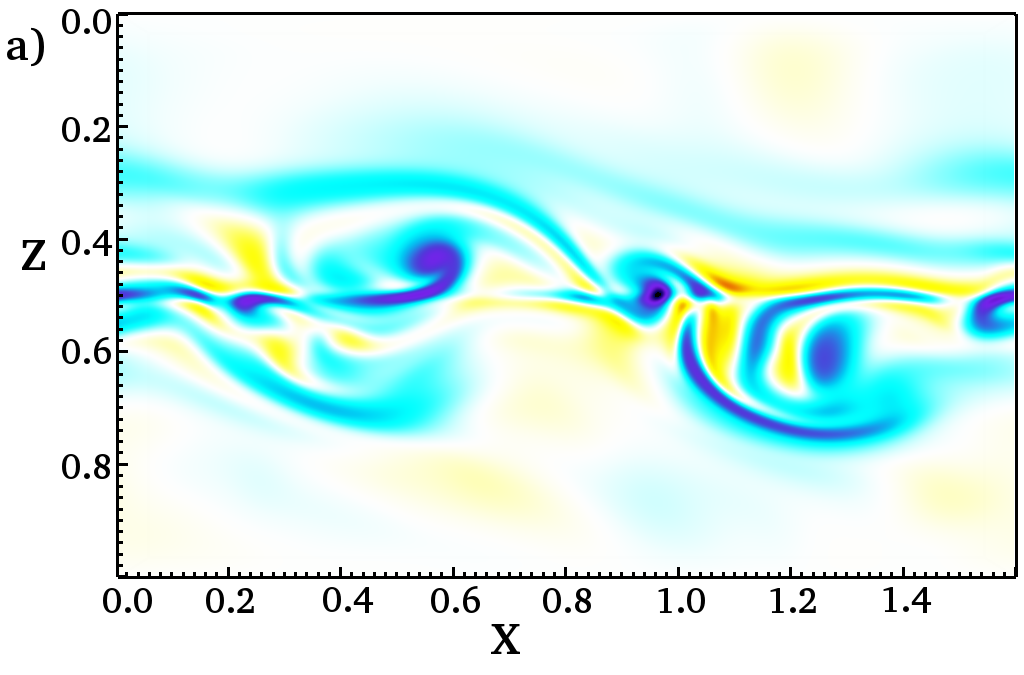}
\includegraphics[width=0.48\textwidth]{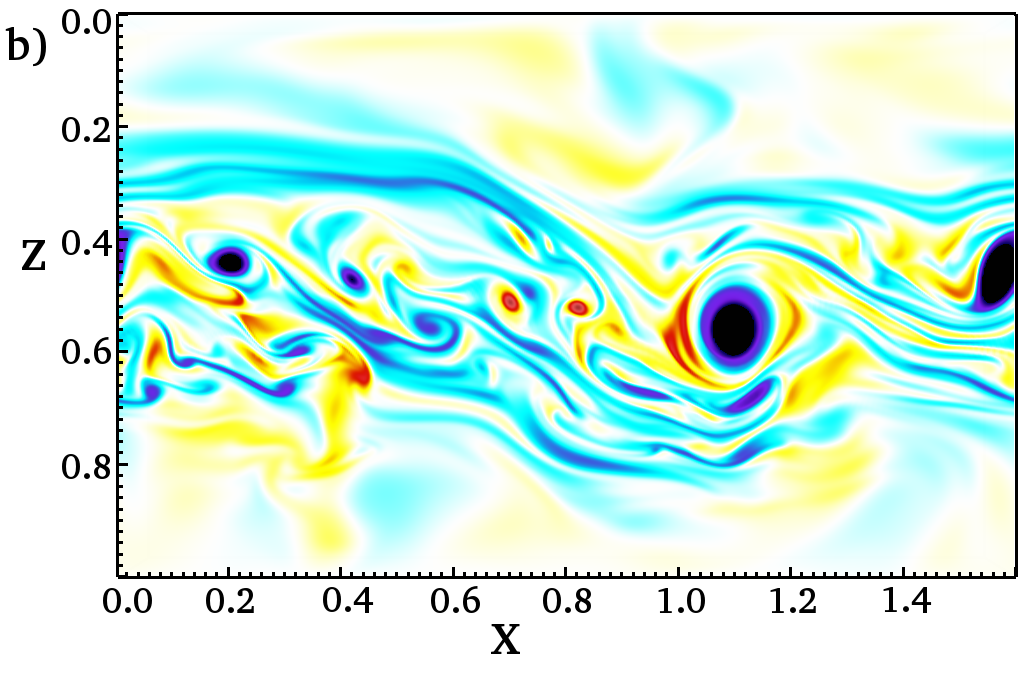}
\includegraphics[width=0.48\textwidth]{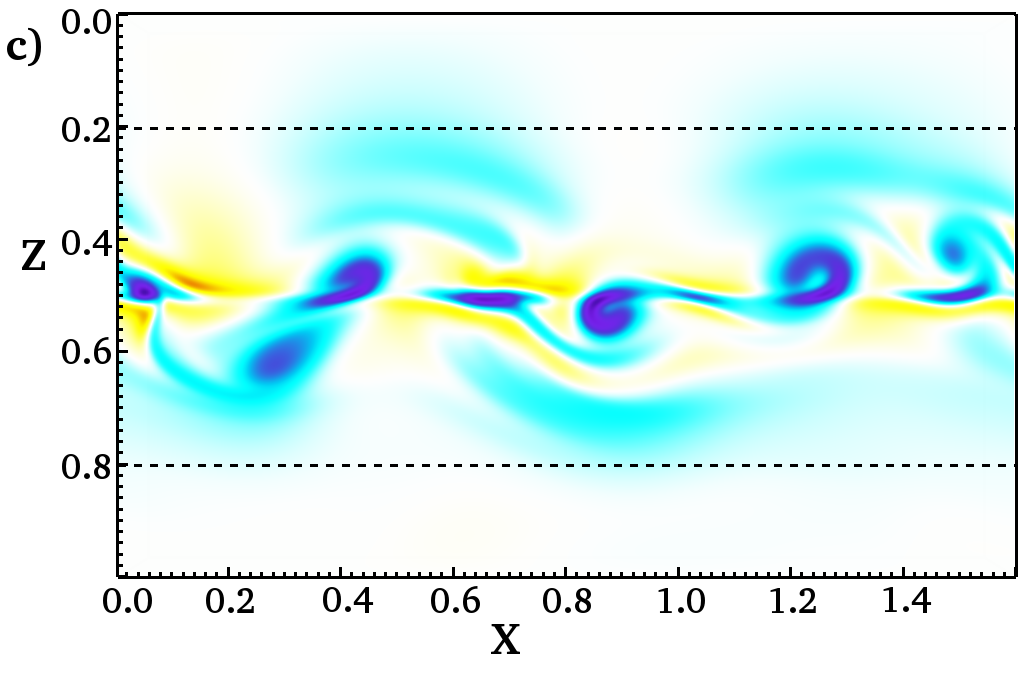}
\includegraphics[width=0.48\textwidth]{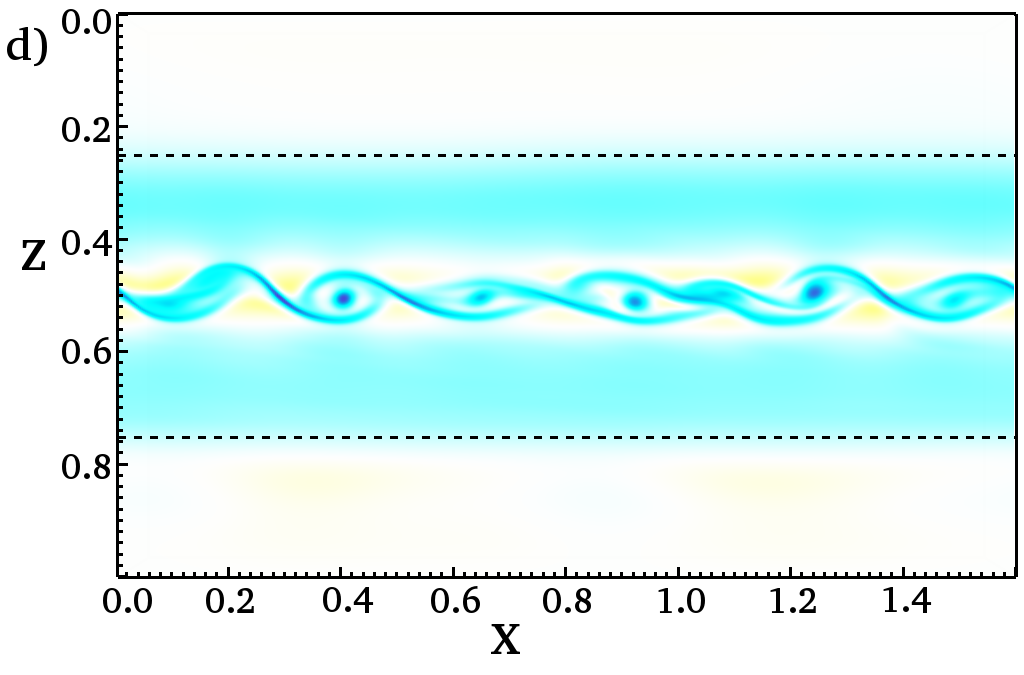}
\includegraphics[width=0.75\textwidth]{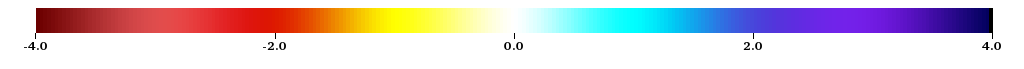}
\caption{Vorticity in the \textit{x-z}-plane for case A and C (see Table \ref{table:two_dim_surv}), where the viscosity is decreasing. At the top (a) and  (b) show case A and C during saturation both at ${t} \approx 93$ and at the bottom the same cases are displayed during the quasi-steady state (c) at ${t} \approx 255$ and (d) at ${t} \approx 500$ (for reference see Fig.~\ref{fig:figure01w}).  For all cases the thermal diffusivity parameter is $C_k = 10^{-4}$. The dotted lines indicate the extent of the turbulent region of the saturated state as obtained from equation (\ref{eq:effectiveL_fit}).}
\label{fig:figure10}
\end{figure*}
In Fig.~\ref{fig:figure10} snapshots for cases A and C are displayed: As the viscosity decreases, vorticity structures are generated on much smaller spatial scales, as expected. However, it also becomes evident that the height of the horizontal layer in which mixing occurs changes only slightly as the viscosity is decreased. In order to estimate the extent of the effective shear region the horizontally averaged velocity in \textit{x}-direction was calculated as
\begin{equation}
\label{eq:u_overbar}
\bar{u}_x(z) = \frac{1}{N_x} \sum_{i=1}^{N_x} u_x(i,z),
\end{equation}
where the overbar denotes that the quantity $u_x$ is horizontally averaged, and $N_x$ is the resolution in \textit{x}-direction. Then, the effective shear width, $L_{eff}$, was obtained by fitting the function 
\begin{equation}
\label{eq:effectiveL_fit}
f(z)= \frac{U_{eff}}{2} \tanh\left(\frac{2}{L_{eff}}{(z-0.5)}\right) 
\end{equation}
to the resulting time averaged $\bar{u}_x(z)$. Since our aim is to approximate the spread, for some cases the middle of the domain was excluded from the fit. To obtain the fit we applied a non-linear least squares method using a trust region algorithm \citep{More_Sorensen_83}  to find $U_{eff}$ and $L_{eff}$. As the goodness of the fit achieves a very small root-mean-squared error of order $10^{-3}$ for all cases, we used the $95\%$ confidence boundaries for the obtained coefficients to estimate the error in $L_{eff}$. Note that, we also tested  a cut-off method as an alternative approach, where we determined the region in which the averaged velocity drops below 95\% of the maximum value of the velocity amplitude, $L_{cut}$. The results are summarised in Table~\ref{table:two_dim_surv} and are qualitatively the same. \\
\begin{figure}
  \vspace*{5pt}
\includegraphics[width=0.49\textwidth]{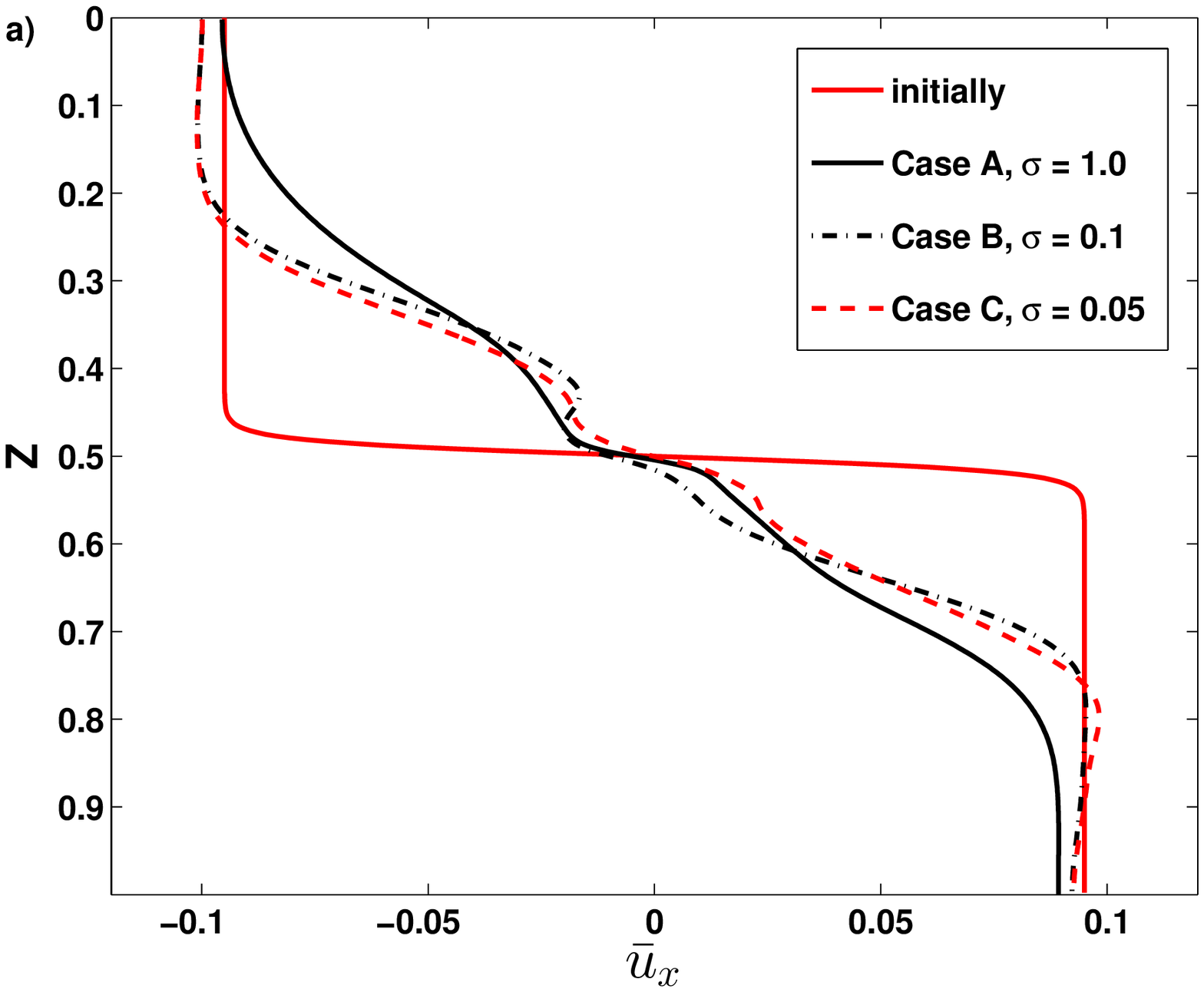}
\includegraphics[width=0.49\textwidth]{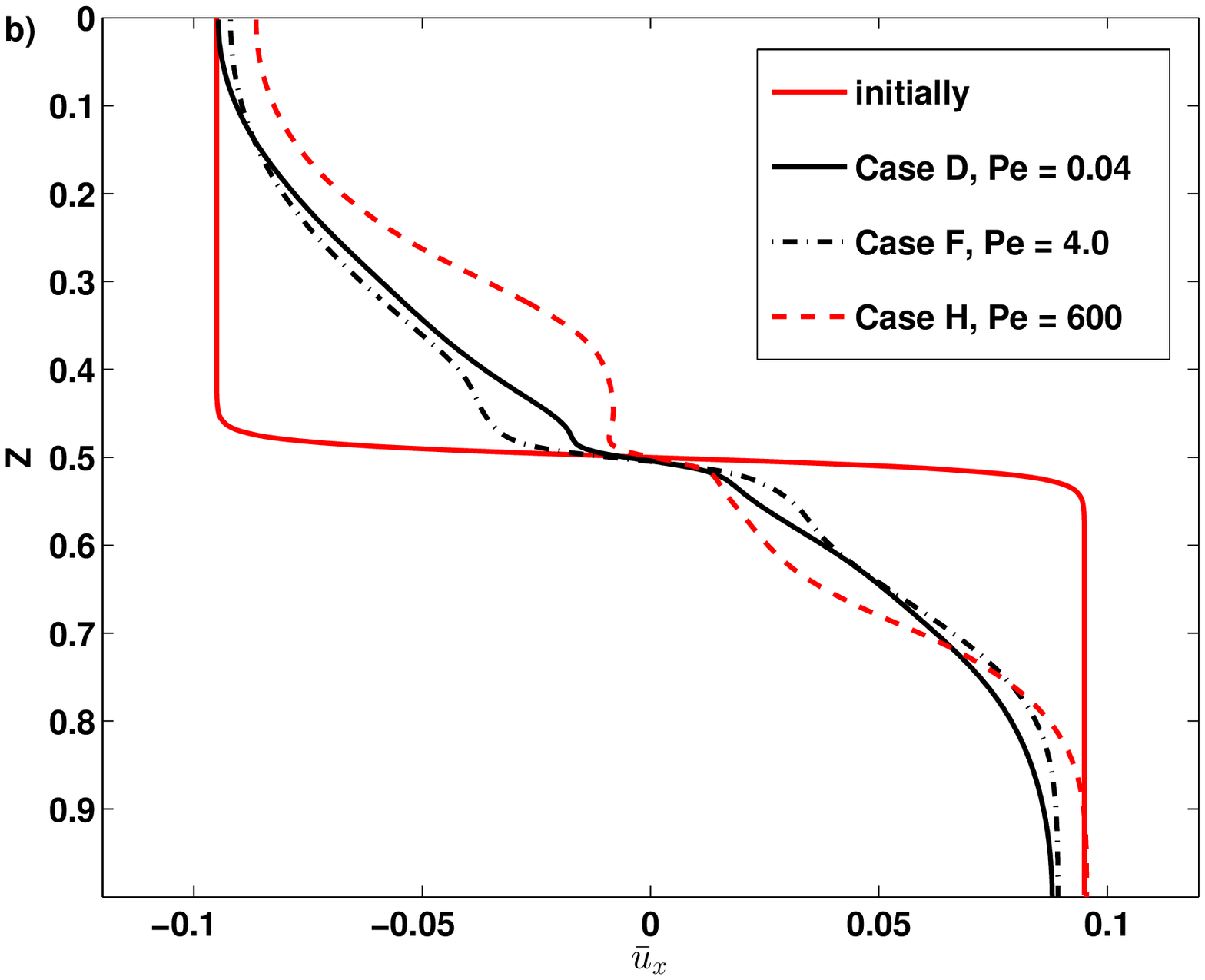}
\caption{The horizontally averaged and time averaged $\bar{u}_x$ profiles are shown for cases A, B, C,  D, F and H.}   
\label{fig:figure11} 
\end{figure}
The averaged velocity profiles, as calculated in equation (\ref{eq:u_overbar}), are displayed in Fig.~\ref{fig:figure11}, where asymmetries can occur due to the fact that we considered  a strongly stratified system. 
For all of the  cases considered the confinement of the effective shear region is due to the dynamics, where the boundaries have only a negligible effect on the form of the horizontally averaged profiles. In Fig.~\ref{fig:figure11}~(a) $\bar{u}_x$ is shown for cases A to C and shows that all of the cases are very similar to each other despite the difference of three orders of magnitude in viscosity. Interestingly, for cases with smaller diffusivities, the form of the averaged velocity profile developes a `staircase like' profile. However, upon checking  the corresponding averaged density profiles, we determined that  no `staircase like' behaviour is present.\\
The effective shear width, $L_{eff}$, for these cases, summarised in Table \ref{table:two_dim_surv}, confirms the previous observation that the vertical extent of the region, where mixing occurs, is barely affected by viscosity:  The effective width decreases only slightly, as viscosity is changed over two orders of magnitude. \\
During the evolution of an unstable shear flow the horizontally averaged profiles for density, temperature and velocity can be  modified. Therefore, the effective minimal $Ri$ number, which we define as
\begin{equation}
\label{eq:effectivRi}
\min Ri_{eff} = \min \left[\frac{-\theta \left(m+1 \right) }{ \gamma \left(\frac{\partial \bar{u}(z)}{\partial z}\right)^2} \left( \frac{\gamma -1}{\bar{\rho}(z)}\frac{\partial \bar{\rho}(z)}{\partial z} - \frac{1}{\bar{T}(z)} \frac{\partial \bar{T}(z)}{\partial z} \right)\right],
\end{equation}
where the overbar denotes horizontally averaged quantities, changes with time. In stratified systems this modification has two sources: The change in the Brunt-V{\"a}is{\"a}l{\"a} frequency, due to changes in the averaged density and temperature profiles, and the change in turnover rates of the shear. In all of the cases we considered, the contributions from density and temperature changes remain negligible compared to the change caused by velocity changes. For all of the cases we considered, the effective Richardson number remains significantly less than the critical Richardson number \citep{2016arXiv161110131W}. Therefore, we conclude that the simple argument that the Kelvin-Helmholtz instability saturates by restoring linear marginal stability \citep[see for example][]{1992A&A...265..115Z, doi:10.1175/2009JPO4153.1, 2014A&A...566A.110P}, is not always valid for complex systems.\\
Understanding the relevant parameters affecting the turbulent characteristics, can provide a comprehensive picture of the possible dynamics in stellar interiors. For the characteristic properties of the turbulent regime the root-mean-square velocity of the perturbations and the typical turbulent length-scales were calculated. The systems that we were considering are stratified such that most quantities will change with depth, \textit{z}, throughout the domain. Therefore, investigating horizontally-averaged profiles varying with depth, before averaging over depth, provides further insight in the dynamics during the saturated regime.
The  horizontal turbulent length-scale of the overall velocity can be defined as
\begin{equation}
\label{eq:turbulent_length}
l_t(z) = 2 \uppi \frac{\int E(k_x,z)/k_x \, dk_x}{ \int E(k_x,z)\, dk_x }, 
\end{equation} 
where $k_x$  is the horizontal wave number. The corresponding energy spectrum $E(k_x,z)$ takes the form 
\begin{equation}
\label{eq:energyspectrum}
E(k_x,z) = \frac{1}{4} \left( \widehat{\mathbf{u}}(k_x,  z) \cdot \widehat{\rho \mathbf{u}}^*(k_x,  z) + \widehat{\rho \mathbf{u}}(k_x,  z)\cdot \widehat{\mathbf{u}}^*(k_x, z) \right),
\end{equation} 
where ${\enspace \widehat{ }\enspace }$ denotes the Fourier transform and the $ ^{*}$ is used to indicate the complex conjugate. In previous studies the vertical scale of the vertical motion was calculated \citep{2016ApJ...821...49G}. 
Due to inherently inhomogeneous nature of the system considered, it is impossible to calculate exactly the same quantity. However, to obtain a comparable quantity we calculated the typical horizontal scale of the vertical motion  by taking only the vertical velocity into account in equation (\ref{eq:energyspectrum}), such that the energy spectrum $E_{w}(k_x,z)$ was obtained. The corresponding turbulent length-scale is then given by 
\begin{equation}
\label{eq:turbulent_vertical_length}
l_w(z) = 2 \uppi \frac{\int E_{w}(k_x,z)/k_x \, dk_x}{ \int E_w(k_x,z)\, dk_x }. 
\end{equation}
However, the resulting typical length-scales, $l_w$, are always smaller than the overall turbulent length-scale, $l_t$, but show the same trends with varied $Re$, $Pe$ and $Ri$  in our investigations.
The root-mean-square of the fluctuating velocity $u_{rms}$(z) we calculated as
\begin{equation}
\label{eq:urms_new}
u_{rms}(z) =  \sum_{x =1}^{N_x}  \sqrt{\left(\mathbf{u}(x,z) - \bar{\mathbf{{u}}}_x(z)\right)^2}/ N_x,
\end{equation}
where $\bar{\mathbf{{u}}}_x(z)$ is the horizontally averaged velocity in $x$-direction as defined in equation (\ref{eq:target_shear}). Here, we averaged over the horizontal layers after the root-mean-square velocity was obtained at each position. The $u_{rms}$ reveals that the actual turbulent region after saturation is more confined as indicated by effective shear width. When the instability is spread the effective shear region is enlarged, but during the saturated regime the outer layers of this region become no longer turbulent. Thus the effective shear region is an upper bound for the turbulent region. 
In addition, we calculated a local turbulent Reynolds number
\begin{equation}
\label{eq:turbulent_reynolds_number}
Re_{t} (z) = \bar{\rho}(z) l_t(z) u_{rms}(z) / (\sigma C_k),
\end{equation}
where $\bar{\rho}(z)$ is the horizontally averaged density. Since our domain it inhomogeneous in the vertical direction the turbulent Reynolds number varies across the domain.\\  
It can be seen in Table \ref{table:two_dim_surv} that $\bar{Re}_t$ increases with decreasing $\sigma$, as expected. 
However, the minimum $l_w$ decreases  from case A to case B but is found to increase  slightly for case C. This indicates that if viscosity is further decreased then the smallest typical length-scales present at the middle of the domain might converge towards a certain value. The $u_{rms}$ increases with increasing viscosity and the maximum $u_{rms}$ at the middle of the domain  as well. Therefore, we conclude that varying $Re$ by changing the Prandtl number does not lead to significant changes of the global characteristics, but affects the typical length-scales of the turbulence as expected.

\begin{figure*}
  \vspace*{5pt}
\includegraphics[width=0.48\textwidth]{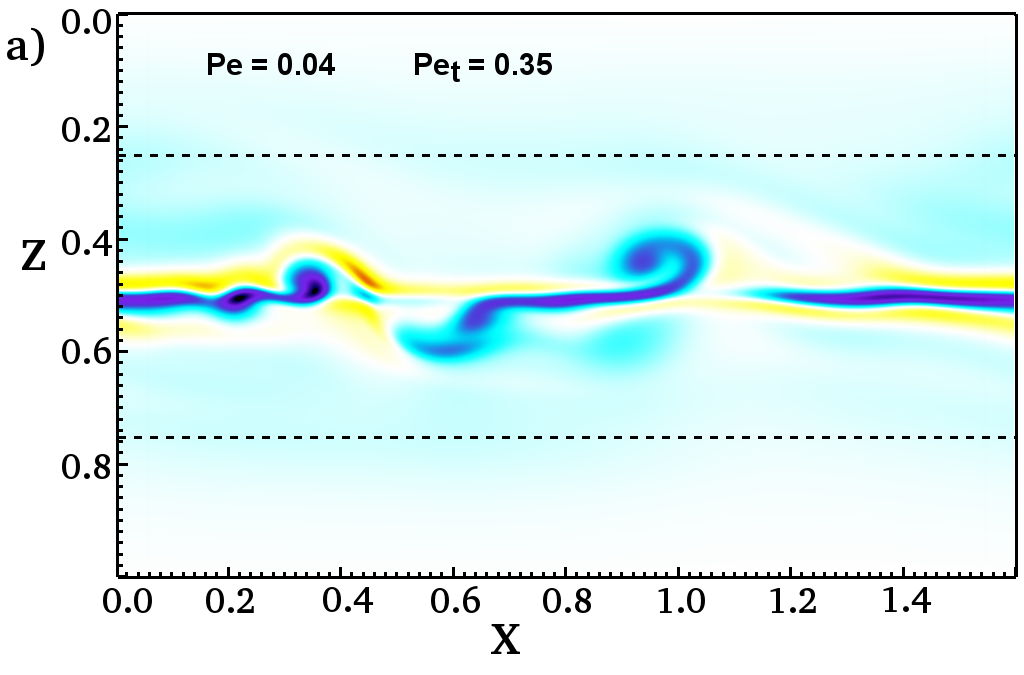}
\includegraphics[width=0.48\textwidth]{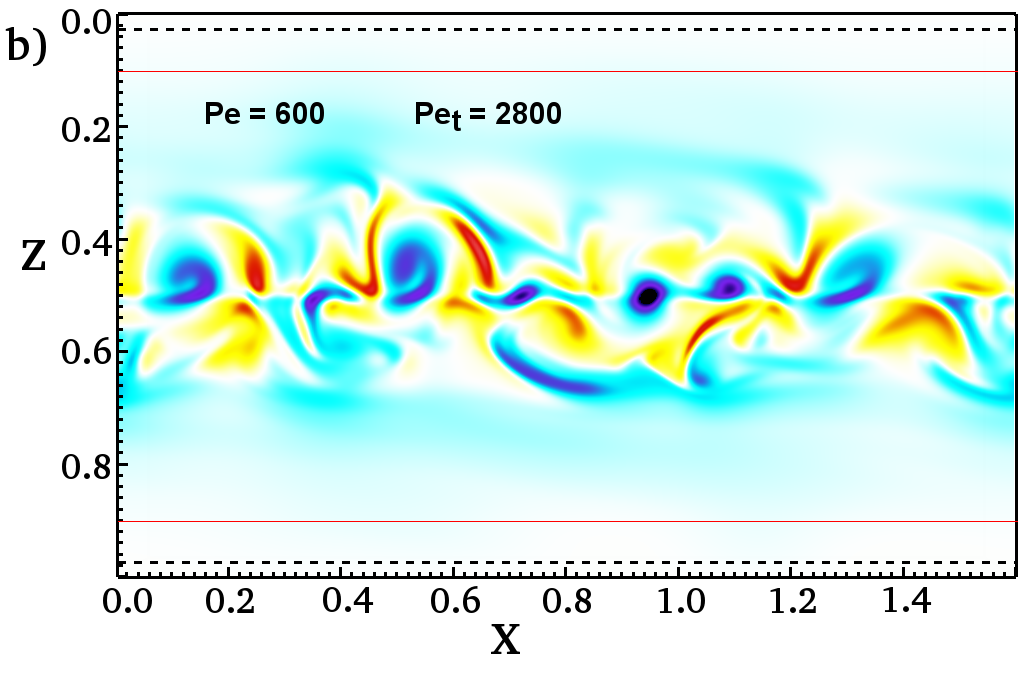}
\includegraphics[width=0.8\textwidth]{./figure2leg}
\includegraphics[width=0.48\textwidth]{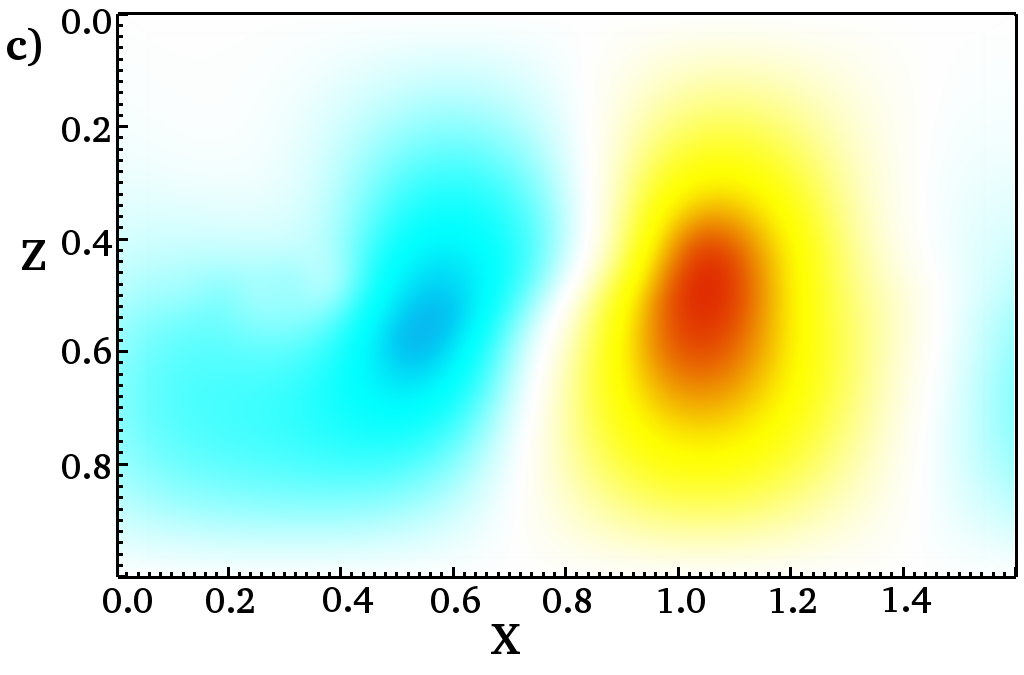}
\includegraphics[width=0.48\textwidth]{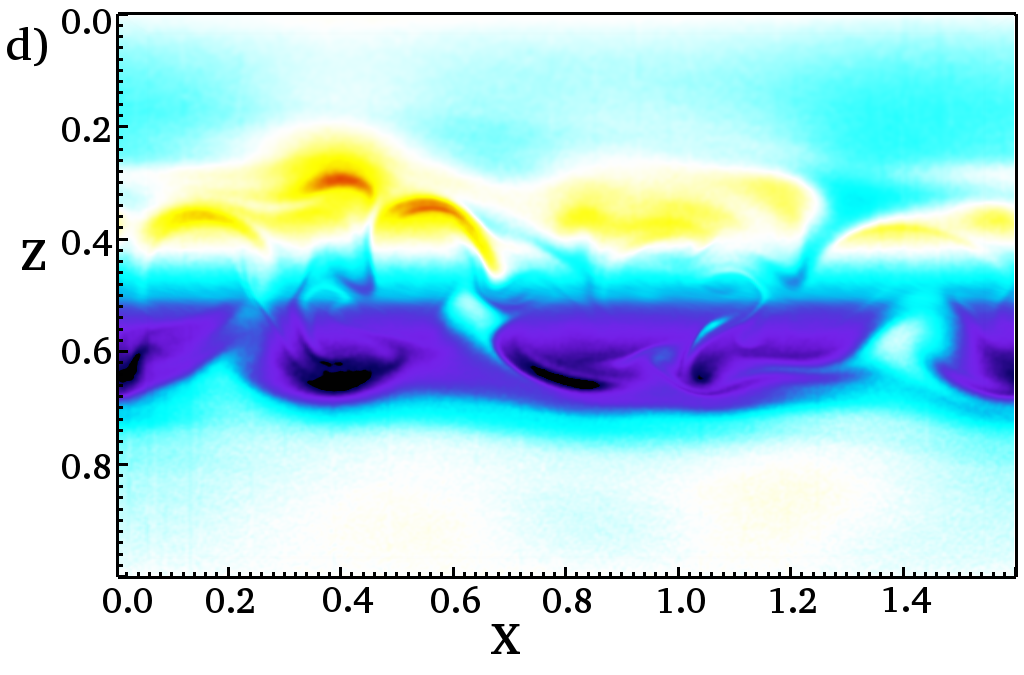}
\includegraphics[width=0.48\textwidth]{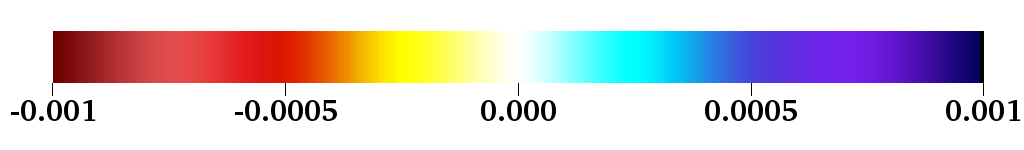}
\includegraphics[width=0.48\textwidth]{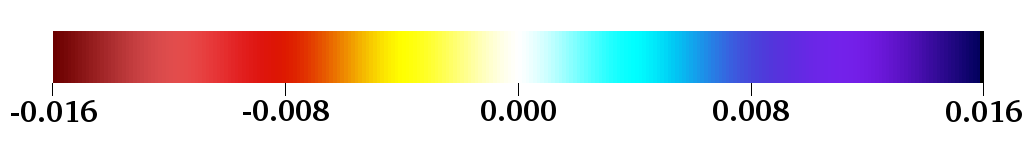}
\caption{The vorticity component perpendicular to the \textit{x-z}-plane (top panel)  and temperature fluctuations around the initial temperature profile (lower panel) for two cases shortly after the system has saturated. Case D is shown in (a) and (c) and case H is displayed in (b) and (d).
 The dotted lines in a) and b) indicate $L_{eff}$ and the red lines in b) show the large error for $L_{eff}$, where the lower bound is indicated. }   
\label{fig:figure_vis_pe} 
\end{figure*}

\subsection{ Varying P\'eclet numbers }
\label{Sec:Vary_peclet_number}
Here the main focus is to investigate different turbulent systems that have different P\'eclet numbers but where both $Re$ and $Ri$ are fixed.  It is important to distinguish between two P\'eclet number limits: In the large P\'eclet number limit  $Pe \gg 1$, which means that the typical time-scale on which advection occurs is shorter than the time-scale on which thermal diffusion acts. The small P\'eclet number limit starts around P\'eclet number of order unity, where the time-scales for advection and diffusion are of the same order, and it continues for all $Pe < 1$.
Moreover, thermal diffusion becomes important in systems where the thermal diffusion time-scale is shorter than the buoyancy time-scale, which  is the case in  a system with $Pe< 1$.  Thus in the small P\'eclet number limit  thermal diffusion weakens the stable stratification, i.e. the system becomes less constrained by buoyancy such that vertical transport is enhanced. We will focus our discussion here on five cases with different $C_k$ and $\sigma$, where $C_k$ was chosen so that the initial P\'eclet number increases by four orders of magnitude over the cases we considered.\\
%
%
In order to examine how the non-linear dynamics change with varying the P\'eclet number we started with a qualitative comparison of the smallest and largest P\'eclet numbers considered for cases D to H. A visualisation of the vorticity after the system has saturated is shown in Fig.~\ref{fig:figure_vis_pe}. For case D, displayed in Fig.~\ref{fig:figure_vis_pe}~(a), with a P\'eclet number of order $10^{-2}$, very strong positive vorticity in a narrow region around the middle plane is present. Here, patches are stretched along the $x$-axis with a few small interruptions of negative vorticity. Positive vorticity regions are stretched outwards from the middle plane at $z=0.5$ and are overturning. 
A significantly different pattern is present in Fig.~\ref{fig:figure_vis_pe}~(b), case H, where the P\'eclet number is of order $10^2$. Here the vorticity amplitude is notably less than in the other case and a vertically extended turbulent region is present. The turbulent region is more isotropic with positive and negative vorticity patches. Furthermore, smaller scale vortices are present in case H compared to case D. From this  we conclude that case D, which is in the small P\'eclet number regime, leads to a different turbulent state, where larger fluid parcels are present compared to case H, which is in the large P\'eclet number regime. This indicates a complex effect of $Pe$ number on the length-scales present in the turbulent regime. Since the P\'eclet number depends on the typical length-scale,  for perturbations on sufficiently short lenght-scales the small P\'eclet number regime is reached. Thus for such perturbations the stabilising effect of the stratification is weaker, which was first noted by \citet{Zahn_1974}, and they can develop. So when decreasing $C_k$ the length-scales that are destabilised become smaller and thus smaller fluid parcels are present. \\
%
Since the  $Pe$ number is varied significantly in cases D to H, another quantity, the temperature fluctuations around the initial background temperature, $\delta T$, is of interest. A visualisation of $\delta T$ for the cases D and H is shown in Fig.~\ref{fig:figure_vis_pe}~(c) and Fig.~\ref{fig:figure_vis_pe}~(d) respectively. The absence of small scale fluctuations in Fig.~\ref{fig:figure_vis_pe}~(c) is a natural consequence of a greater thermal diffusivity.\\ 
Note, when  increasing the $Pe$ number, from D to H, the extent of the turbulent layer after saturation looks visually larger. This is confirmed by the effective shear width, $L_{eff}$ in Table \ref{table:two_dim_surv}. Since the dynamical viscosity is fixed, these effects result solely from different $Pe$. Therefore,  the P\'eclet number, which is associated with the thermal diffusion,  plays an important role in the non-linear dynamics and in particular on the vertical spread of the shear induced turbulence. \\
We find that in the limit of large P\'eclet numbers the effective spread increases with increasing  P\'eclet numbers as the effective shear width, $L_{eff}$, becomes larger for cases F to H as shown in Table \ref{table:two_dim_surv}. This shows that a decreased $Pe$ number significantly damps the spread of perturbations. In the large $Pe$ regime the temperature of a fluid parcel will adjust faster to the surrounding temperature when $Pe$ is decreased. Therefore, part of the kinetic energy is irreversibly converted into internal energy quicker and so the further propagation of the fluid parcels is hindered. However, in the limit of small P\'eclet numbers the opposite trend is observed, where the effective shear width decreases with increasing $Pe$. This suggests a complex interplay between the effect of thermal diffusivity and energy contained in the system, which we investigate in the next subsection. \\
We compared the typical turbulent length-scale, $l_w$, and the root-mean-square of the velocity perturbations, $u_{rms}$, for different initial P\'eclet numbers (cases D to H in Table \ref{table:two_dim_surv}). These cases reveal that the $Pe$ number significantly affects the turbulence scale $l_w$, where the smallest $l_w$ is found for the case with $Pe = 60$.
Furthermore, the maximum root-mean-square of the velocity perturbations (see Table~\ref{table:two_dim_surv})  reduces with increasing thermal diffusion in the limit of large P\'eclet numbers. This trend is consistent with the recent Boussinesq investigations of the low P\'eclet number regime \citep{2016ApJ...821...49G} where the Richardson number was varied. However, in our setup the observed turbulent length-scales always exceed the minimal pressure scale-height, $H_{min}$,  by at least a factor of two, whereas for some cases the maximum  pressure scale-height, $H_{max}$ is slightly larger than the turbulent length-scale. So  that the Boussinesq approximation is not valid for our system.\\
%
%
%
%
%
%
%
\subsection{Varying the Richardson number}
\label{sec:Ri_Re_varying}
\begin{figure}
  \vspace*{5pt}
\includegraphics[width=0.41\textwidth]{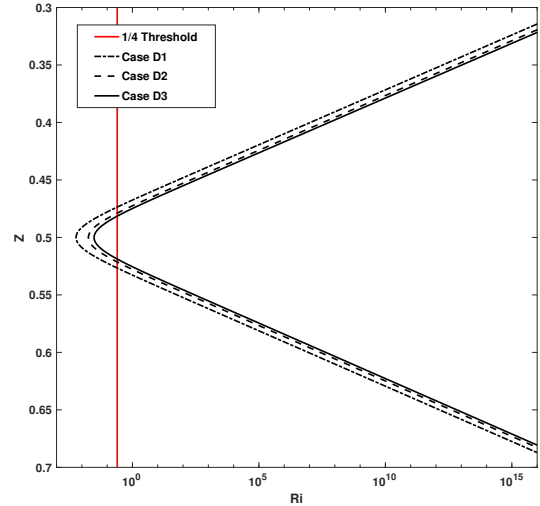}
\caption{$Ri$ with depth for the cases D1, D2 and D3.}   
\label{fig:figure_3.3RiPe} 
\end{figure}
To understand how the energy provided by the background flow will affect the saturated regime, the Richardson number was varied independently of all other parameters. This was achieved by changing $\theta$ separately for a low $Pe$ number (cases D1, D2, D3 ), an intermediate $Pe$ number (cases FF1, FF2, FF3), and for a large $Pe$ number (cases G1, G2, G3), which are summarised  in Table~\ref{table:two_dim_surv}. For $Pe = 0.04$  and $Pe = 6$ we changed the Richardson number by one order of magnitude via varying the thermal stratification between $\theta = 0.8$ and $\theta = 6$. Fig.~\ref{fig:figure_3.3RiPe} shows the $Ri$ across the middle of the domain for the cases D1 to D3. The region where this number remains less than 1/4 decreases slightly with increasing $Ri$. As a result the effective shear width significantly decreases with increasing Richardson number for the cases D1 to D3 and FF1 to FF3. The Richardson number is proportional to the ratio of the potential energy that is needed to overcome the stabilising stratification and the available kinetic energy of the background flow. Since a perturbation loses its initial energy when moving vertically, in the ideal case it will continue to spread as long as it has more energy than needed to overcome the stratification. Therefore, the vertical extent of the turbulent region increases with decreasing Richardson numbers. However, in the limit of large $Pe$ the opposite is observed, and so the effective shear widths increases with increasing $Ri$, which indicates a more complex effect. Therefore, we conclude that there are two parameters affecting the extent of the effective shear region generated by an unstable shear flow in a stratified system. The first is the Richardson number, since it provides information on the ratio between available kinetic energy and the potential energy. The second parameter that controls the extent of the effective shear region is the P\'eclet number, which affects the vertical motion of fluid parcels.\\
The product of the Richardson number and the P\'eclet number, $Ri Pe$, has been used as an input parameter in Boussinesq calculations  \citep{2014A&A...566A.110P, 2016ApJ...821...49G} and so it is natural to consider if this product can be used to quantify the system in these  compressible calculations. Therefore, we now focus on comparing the typical length-scale and root-mean-square of the velocities obtained in the cases D1 to G3 with results obtained in \cite{2016ApJ...821...49G}. For the small $Pe$ regime the decrease of the typical length-scale with increasing $Ri Pe$, is recovered. However, the root-mean-square velocities in our calculations decrease with increasing $Ri Pe$, whereas in \cite{2016ApJ...821...49G} they increase with increasing $RiPe$. Interestingly, in the large $Pe$ regime the typical length-scale increases with $RiPe$, if the P\'eclet number is increased, but decreases if the $Ri$ number is increased. This result suggests that in stratified systems, and also in the large $Pe$ regime the product of $Ri$ and $Pe$ can not be used to characterise the saturated dynamics of the system. It is necessary to consider each dimensionless number separately.
%
%

\subsection{Secular Instability}
\label{Sec:Secular_Instability}
\begin{table*}
\setlength\tabcolsep{5pt}
\centering
\caption{For the `secular' instability cases O to R the temperature gradient, $\theta = 2.0$, and the polytropic index, $m=3.0$, are fixed, but the initial $Ri$ number changes from 0.4 for cases O to Q to 1.0 for case U.  For all cases the pressure scale-height is the same, where $H_{max} = 0.4$ and $H_{min}= 0.15$. The initial P\'eclet number,$Pe$, Reynolds number, $Re$, and minimum Richardson number, $Ri$, are listed. The effective shear width is calculated after saturation and  $\bar{Re_t}$,  and $\bar{l_w}$  are averaged  over the whole domain. Additionally, we provide a shear width, $L_{cut}$ obtained by a cut-off method at 95\%. }
\begin{tabular}{c c c c c c c c c c c c c c c   }
\hline
\hline
\multicolumn{6}{c}{Secular Instability} & \multicolumn{6}{c}{Resolution $N_x = 512$, $N_z = 480$  }\\
\hline
\textbf{ Case:} &  $\sigma$ & $C_k$  &  $U_0$  &$L_u$ & $Pe$ &$Re$ &  $Ri$  & $L_{eff}$ & $L_{cut}$ & $\bar{Re_t}$  & $\bar{u_{rms}}$ & $\max u_{rms} $ &$\bar{l_w}$ &  $\min(l_w) $ \\
\hline
O& 0.001 &0.05  &  0.1 & 0.029  & 0.06 & $5.7 \times 10^1 $ & 0.4  & $0.076 \pm 0.002$ & 0.20 &$4.8 \times 10^1$ & $8.5 \times 10^{-04}$ &$ 6.7 \times 10^{-3}$ &0.58 & 0.20 \\
P & 0.0006 & 0.05& 0.1& 0.029 & 0.06& $9.5 \times 10^1 $ & 0.4  & $0.094 \pm 0.002$  & 0.24 &$1.6 \times 10^2$  & $1.1 \times 10^{-03}$ & $ 7.4 \times 10^{-3}$  &1.16 & 0.26 \\
Q & 0.0002 & 0.05 &  0.1& 0.029 & 0.06& $2.9 \times 10^2 $ & 0.4  & $0.097 \pm 0.001$  & 0.25 &$8.1 \times 10^2$  & $1.2 \times 10^{-03}$ & $ 6.5 \times 10^{-3}$  &1.12 & 0.36 \\
R & 0.0002 & 0.05 &    0.07& 0.033 & 0.05&  $2.3 \times 10^2 $ & 1.0 & $0.063 \pm 0.001$& 0.17 &$7.8 \times 10^2$  & $9.8 \times 10^{-04}$ & $ 6.3 \times 10^{-3}$  &1.4 & 0.82 \\

\end{tabular}
\label{table:secular_Instability}
\end{table*}
\begin{figure}
  \vspace*{5pt}
\includegraphics[width=0.45\textwidth]{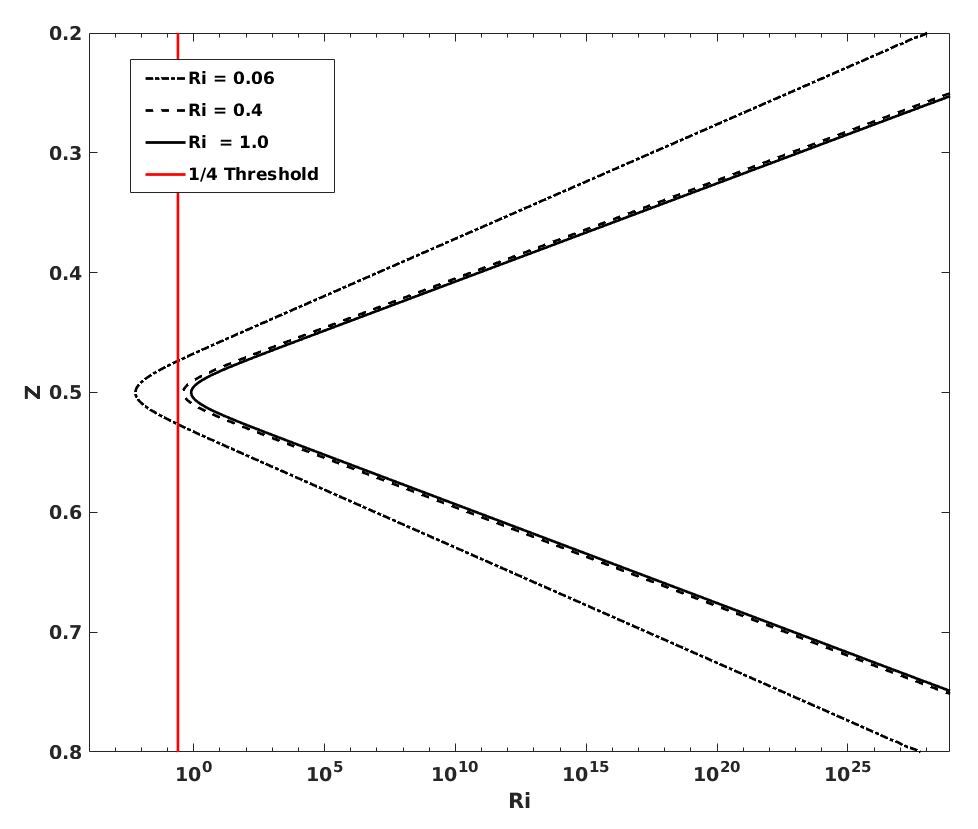}
\caption{Richardson number with depth around the middle of the domain for the secular instability cases and the most of the cases from Table~\ref{table:two_dim_surv}.}   
\label{fig:figure_diffRi} 
\end{figure}
In the previous section we focused only on classical shear flow instabilities that can be present within the stellar interiors where Richardson numbers become smaller than 1/4.  However, another well-known type of shear instability can occur even when the Richardson number is significantly greater than 1/4 \citep{Dudis_1974, Zahn_1974, Garaud10.10631.4928164}, but only in the low P\'eclet number regime.\\
Low P\'eclet numbers are most likely to be present in upper regions of massive stars, where differential rotation is present and `secular' shear instabilities can develop.   Previous investigations have considered diffusive systems either by using the Boussinesq approximation \citep{1999AA...349.1027L, 2013A&A...551L...3P, Garaud10.10631.4928164} or restrict the study to linear stability analysis when employing the fully compressible equations \citep{2015AandAWitzke}.  In order to investigate the evolution that occurs during the saturated phase, non-linear studies are required. However, it is extremely difficult to consider small P\'eclet number regimes using the full set of equations as the time-stepping is restricted by stability constraints related to the diffusion time. Therefore, it has previously been impossible to numerically investigate small P\'eclet number shear flows without using approximations \citep[as in][]{2013A&A...551L...3P}. However, we show that it is possible to calculate low P\'eclet and large Richardson number cases in a  two-dimensional domain. Here we present some calculations,  with a spatial resolution of $480 \times 512$ to investigate the differences between the classical and `secular' instabilities during the saturated phase.\\
To ensure that the classical KH instability is not triggered, we set $Ri =0.4$, which  was achieved by taking $\theta = 2$, $m=3.0$, $U_0 = 0.1$ and $1/L_u =70$. Taking $C_k = 0.05$ leads to an initial P\'eclet number of order $10^{-2}$, such that a secular instability can develop due to the destabilising effect of low P\'eclet numbers. Fig.~\ref{fig:figure_diffRi} shows how the $Ri$ number changes in the middle of the domain comparing the profile to the small $Ri=0.006$ cases investigated above. For cases O, P and Q the Reynolds number was varied via the Prandtl number  in order to investigate its effect on the turbulent length-scale. In order to study how different $Ri$ affect the dynamics of a secular instability,  for case R the Richardson number was increased to $Ri =1.0$. This was achieved by changing $U_0 = 0.07$ and $1/L_u =60$, and the dynamical viscosity was fixed to be $10^{-5}$. We compared the growth rate and the most unstable mode predicted by means of a linear stability analysis as used in \citet{2015AandAWitzke} to that found in all cases of the secular instability. Furthermore, we  checked that the instability is a consequence of the destabilising mechanism at low P\'eclet numbers, by conducting test cases with the same dynamical viscosity as used in cases O to R but $C_k = 0.0002$. This result in an initial $Pe \approx 12$ and we find that for both Ri numbers the system remains stable and the initial perturbations decay.\\
%
%
\begin{figure}
\centering
  \vspace*{5pt}
\includegraphics[width=0.07\textwidth]{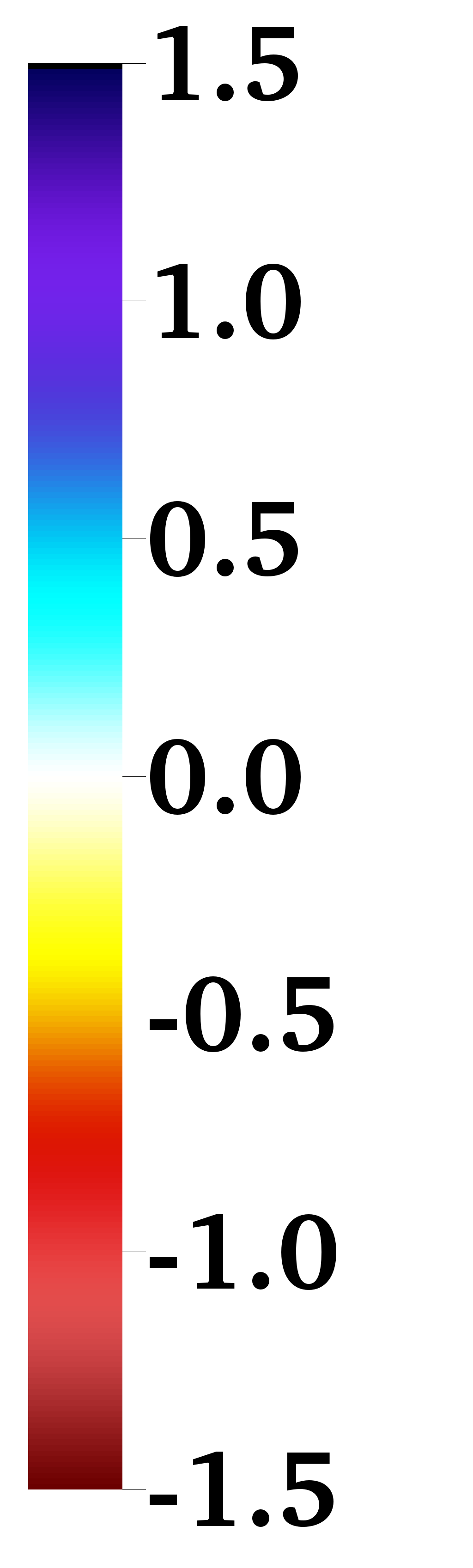}
\includegraphics[width=0.4\textwidth]{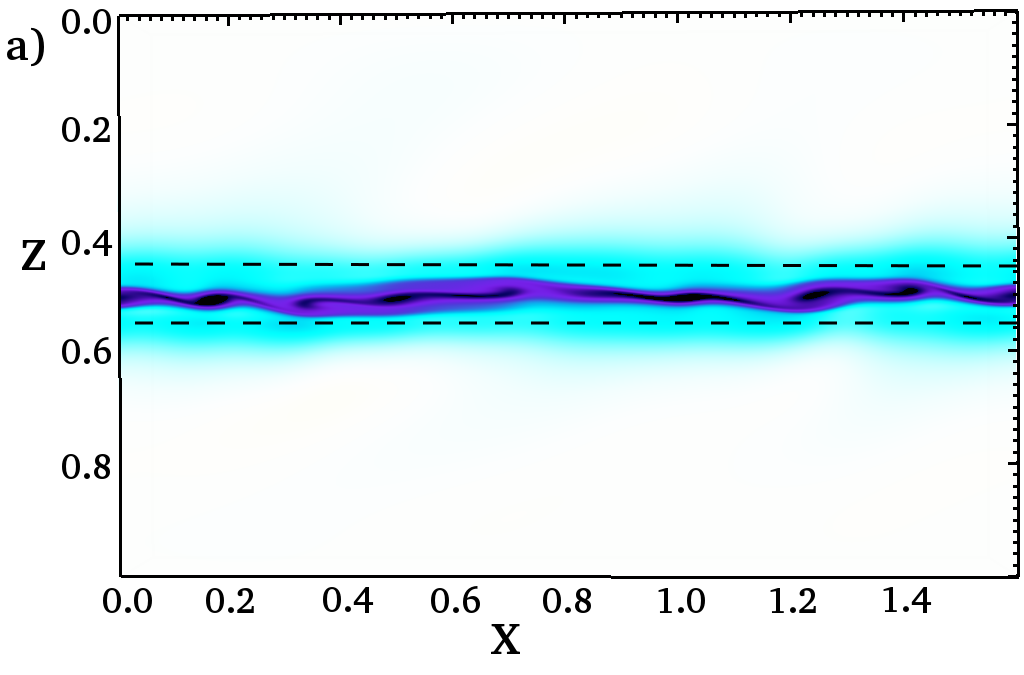}

\includegraphics[width=0.07\textwidth]{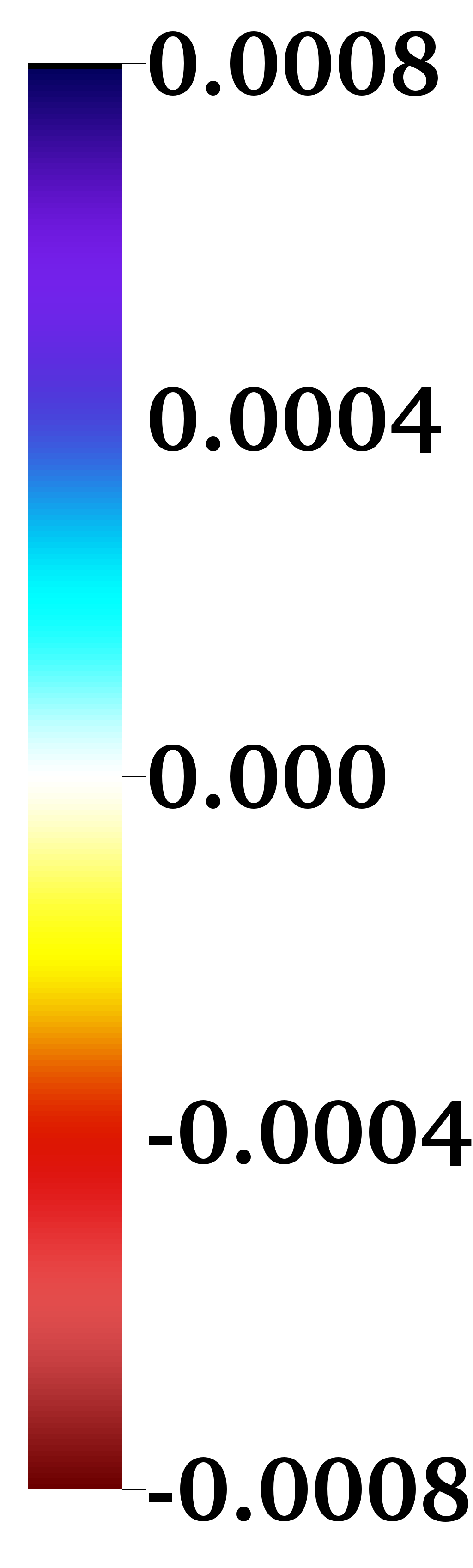}
\includegraphics[width=0.4\textwidth]{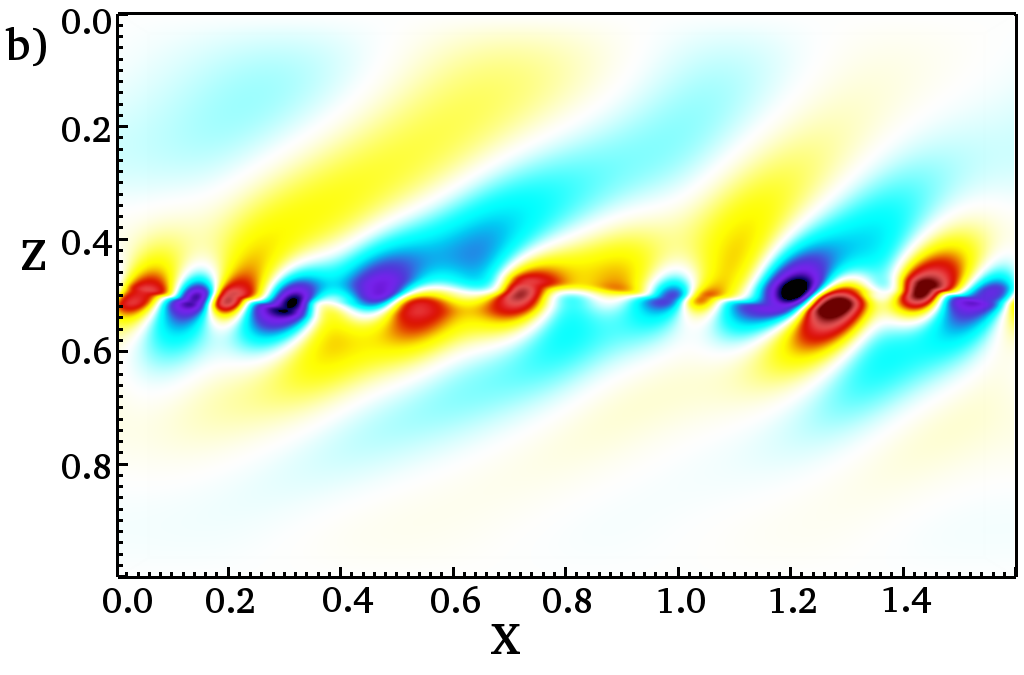}

\includegraphics[width=0.07\textwidth]{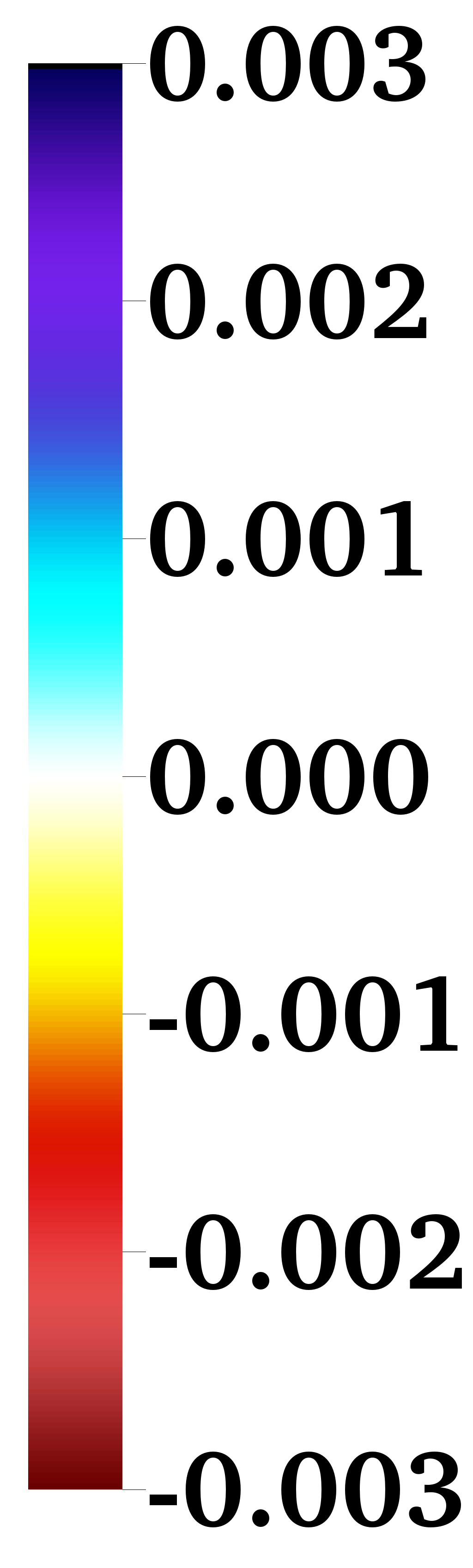}
\includegraphics[width=0.4\textwidth]{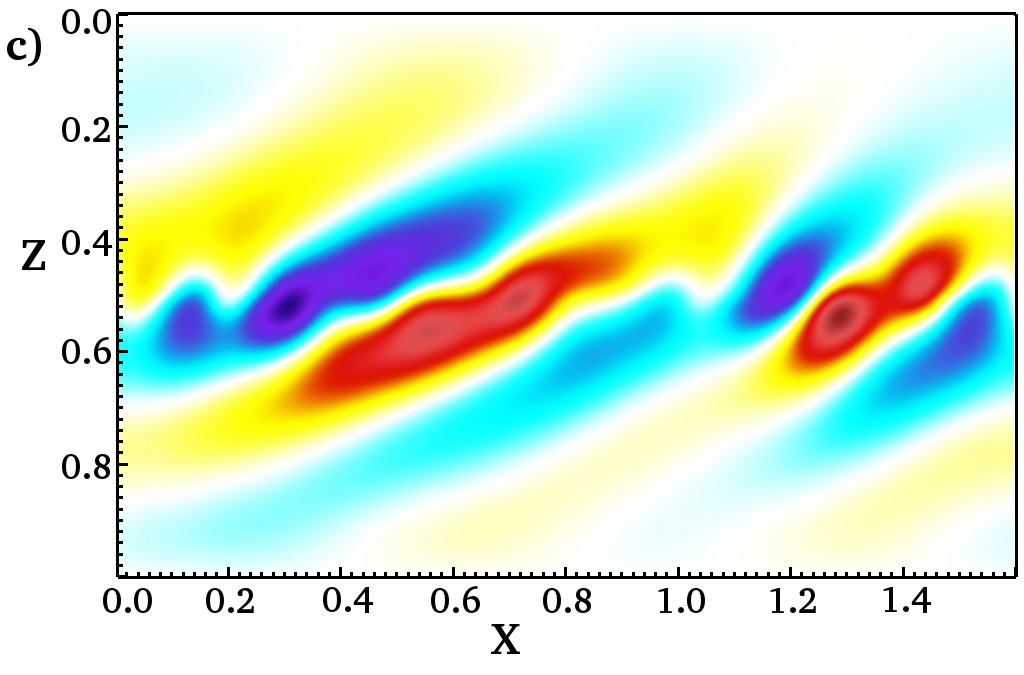}

\caption{The vorticity component perpendicular to the \textit{x-z}-plane, the vertical velocity, $w$, and temperature fluctuations around the initial temperature profile for case Q long after saturation. In (a) the vorticity is shown and the dotted line indicates the extent of the effective shear, $L_{eff}$, in (b) the vertical velocity is shown, and (c) shows the temperature fluctuations.}   
\label{fig:figure_05} 
\end{figure}
When the instability starts to saturate in any of the cases O to R we observe very little overturning, which is unlikely to develop into a turbulent regime at least for the Reynolds numbers considered here (see Fig.~\ref{fig:figure_05}). For all cases the vertical spread of the flow is significantly smaller than observed in the previous cases (see Fig.~\ref{fig:figure11}~(c) and Table~\ref{table:secular_Instability}). This is because the Richardson numbers are very large and the available kinetic energy for the perturbations is decreased.  We find for all cases here that the vertical spread of the temperature perturbations is slightly greater than it is for the vertical velocity during the saturation phase, which can be seen from  Fig.~\ref{fig:figure_05} (b) and Fig.~\ref{fig:figure_05} (c). A similar trend appears for a classical shear instability in the low P\'eclet number regime as can be seen in Fig \ref{fig:figure_vis_pe}. While there is some similarity in the trend in terms of the spread, there is a marked difference in the patterns observed  in the saturated state between the secular and the classical instabilities.
Fig.~\ref{fig:figure_05}~(b) shows that regions of upward and downward motion are stretched along the horizontal direction. Similar pattern of the negative and positive temperature fluctuations is present in Fig.~\ref{fig:figure_05}~(c), where layers are formed. Comparing what is seen in Fig.~\ref{fig:figure_05}~(c) to the  temperature fluctuations present in case D (see Fig.~\ref{fig:figure10}~(d)), where the P\'eclet number is of the same order, we see that for the classical instability no layering occurs. \\
Turning to the characteristic length-scale during the saturated regime we find that, from the data in Table \ref{table:secular_Instability}, decreasing the viscosity results in greater typical turbulent length-scales. The opposite trend was observed in the previous study of an unstable system at large P\'eclet number. The trend for the `secular' instability cases can be explained by the peculiar pattern observed for the vertical velocity and temperature fluctuations as shown in Fig.~\ref{fig:figure_05}. Since the fluctuations are sheared out the typical length-scale between the up and down motions is increased.  When the viscosity is decreased it becomes easier for the horizontal movement  of the background flow to elongate the fluctuation pattern even more, such that the typical length increases. However, such a pattern does not transit to developed turbulence for the cases considered. In a test case with $Ri = 0.1$, a similar layering is observed when the P\'eclet number dropped below unity. However, case D, which has a $Pe = 0.04$ and a $Ri = 0.006$, does not show any similar layering. Therefore, we conclude that such layering can develop in systems that are not very far from the stability threshold. Such a behaviour is not only a feature of the secular instability, but can be present in a system with Richardson number close to the stability threshold, but with a sufficiently small P\'eclet number. \\
To summarize, we find that the `secular' shear flow instability in a fully compressible, stratified fluid, shows the expected trend for the spread of the instability. Although the cases studied do not become turbulent, a turbulent regime will be eventually reached when considering larger Reynolds numbers.  We observe a difference in the trend for typical length-scales during the saturated regime, when compared to cases where a classical KH instability was triggered.  Moreover, a peculiar pattern for the temperature fluctuations and vertical velocity fluctuations are found to be present. 

%
\subsection{Saturated regime using three-dimensional calculations}
\label{sec:three_dim_long_time_study} 
\begin{figure*}
  \vspace*{5pt}
\includegraphics[width=0.49\textwidth]{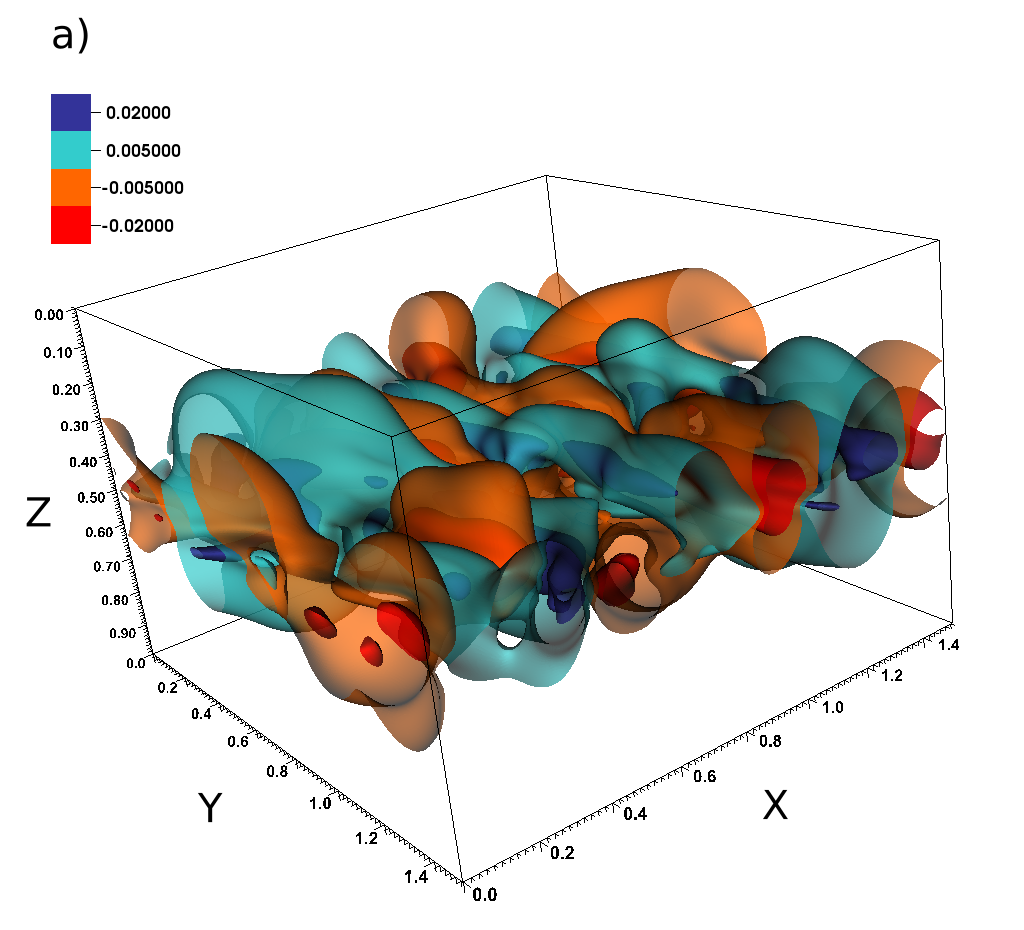}
\includegraphics[width=0.49\textwidth]{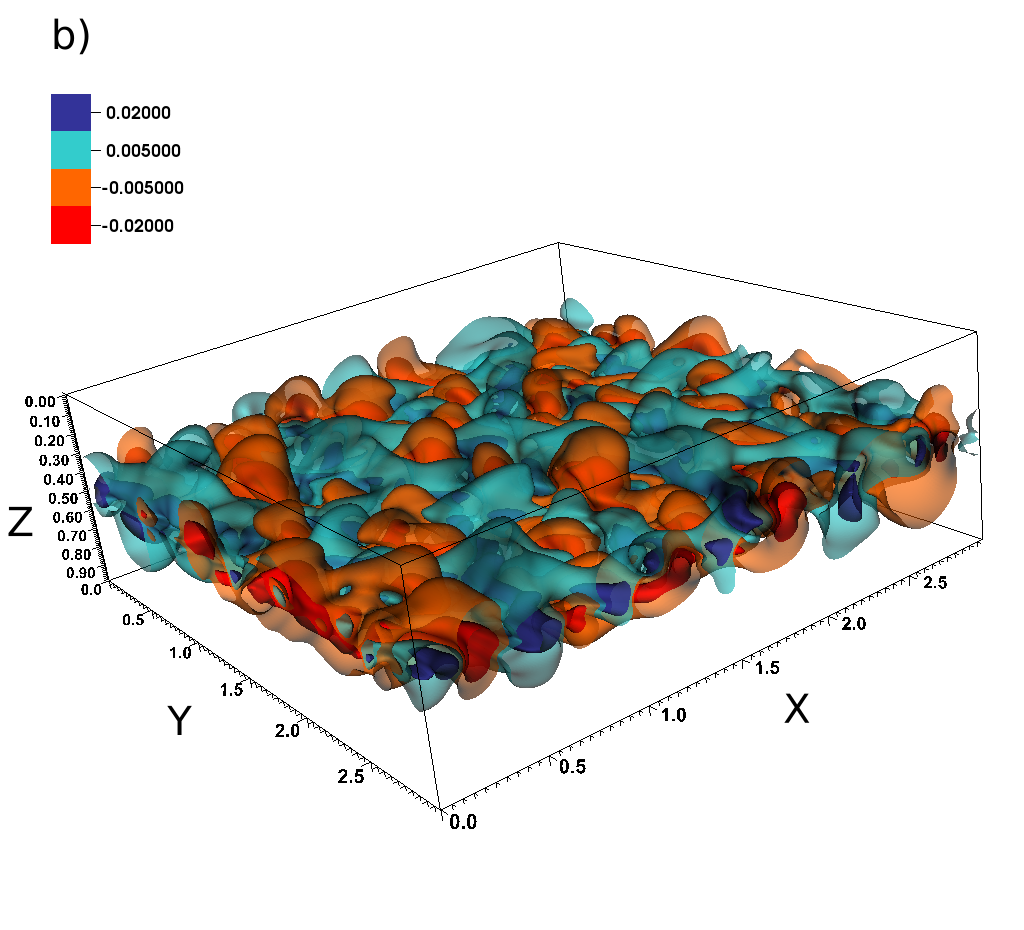}
\caption{The isosurfaces of constant vertical velocity, $w$, for four different velocity values in three-dimensional calculations. (a) case I at roughly $t \approx 200$. (b) case J at $t \approx 110$. }   
\label{fig:figure_3dim} 
\end{figure*}
%
Two-dimensional calculations may alter the non-linear dynamics from what could occur in three dimensions, since vortex stretching is suppressed. Therefore, it is crucial to investigate two representative cases in three dimensions to confirm the non-linear dynamics. Therefore, here we will discuss cases where the full set of dimensionless, three-dimensional, differential equations is used. An additional horizontal $y$-direction prependicular to the $z$-$x$-plane is considered, which is periodic and its length is normalised by the depth $d$.\\
We chose two different P\'eclet number regimes, because both small and large P\'eclet number regimes can occur in stellar interiors depending on the region considered and the type of star. Here, case I represents the large P\'eclet number regime, $Pe > 1$ and case J the low P\'eclet number regime with $Pe < 1$.  The calculations were evolved over a sufficient fraction of the largest thermal diffusion time-scale in the system, which is given by the domain depth squared divided by the thermal diffusivity, $t_{thermal} = 1/C_k$, where $1$ is the non-dimensional  depth of the domain. The low $Pe$ case was even evolved for several thermal times, $t_{thermal}$.  The spatial resolution for these calculations is $N_x = 256$, $ N_y = 256$ and $ N_z = 320$. The spatial extent of the horizontal dimensions needs to be taken larger for case J, because in a smaller box the secondary instability that  propagates in the $y$-direction is suppressed. The exact parameters for these  cases are summarized in Table \ref{table:two_dim_surv}, where the corresponding two-dimensional cases K and L have the same parameters and resolution as their corresponding three-dimensional case. Note, that this comparison study was performed at a greater viscosity, $\sigma C_k$,  than all other cases due to computational cost.     \\
Here, we compared the characteristics of the three-dimensional calculations to two-dimensional calculations. Furthermore, direct comparisons of global properties obtained in the saturated regime for both the small and large P\'eclet number regime calculations were performed. \\
Fig.~\ref{fig:figure_3dim} shows contour plots of the vertical velocity, $w$, for case I and case J. These plots were taken a long time  after the system has saturated for both cases. Visually the three-dimensional calculations show similar dynamics:
The turbulent regions have approximately the same vertical extent, as well as the overturning regions, for both considered cases. This is explained by the fact that the initial P\'eclet numbers of the two cases are close to the threshold between small and large P\'eclet number regime, such that the dynamics in both regimes are similar. Similar dynamics are also observed in two-dimensional calculations for the small P\'eclet number regime. However, here in the three-dimensional calculations, secondary instabilities lead to turbulent motions in the $y$-direction. Fig.~\ref{fig:figure_3dim} clearly shows a well developed turbulence in both horizontal directions.\\
When comparing the three-dimensional cases to their corresponding two-dimensional cases K and L, the effective shear width $L_{eff}$ is similar for the large $Pe$ calculations. However, in the small $Pe$ regime the three-dimensional case has a reduced shear width $L_{eff} = 0.40$ compared to the corresponding two-dimensional case, where it is $L_{eff}=0.48$. Although the values differ between the two- and three-dimensional cases, it is important to note that a similar increase in the spread for larger $Pe$ persists. The discrepancies in the confinement in three-dimensional calculations can be explained by the different non-linear dynamics where, for example, a secondary instability can evolve perpendicular to the x-direction. \\
\begin{figure}
  \vspace*{5pt}
\includegraphics[width=0.48\textwidth]{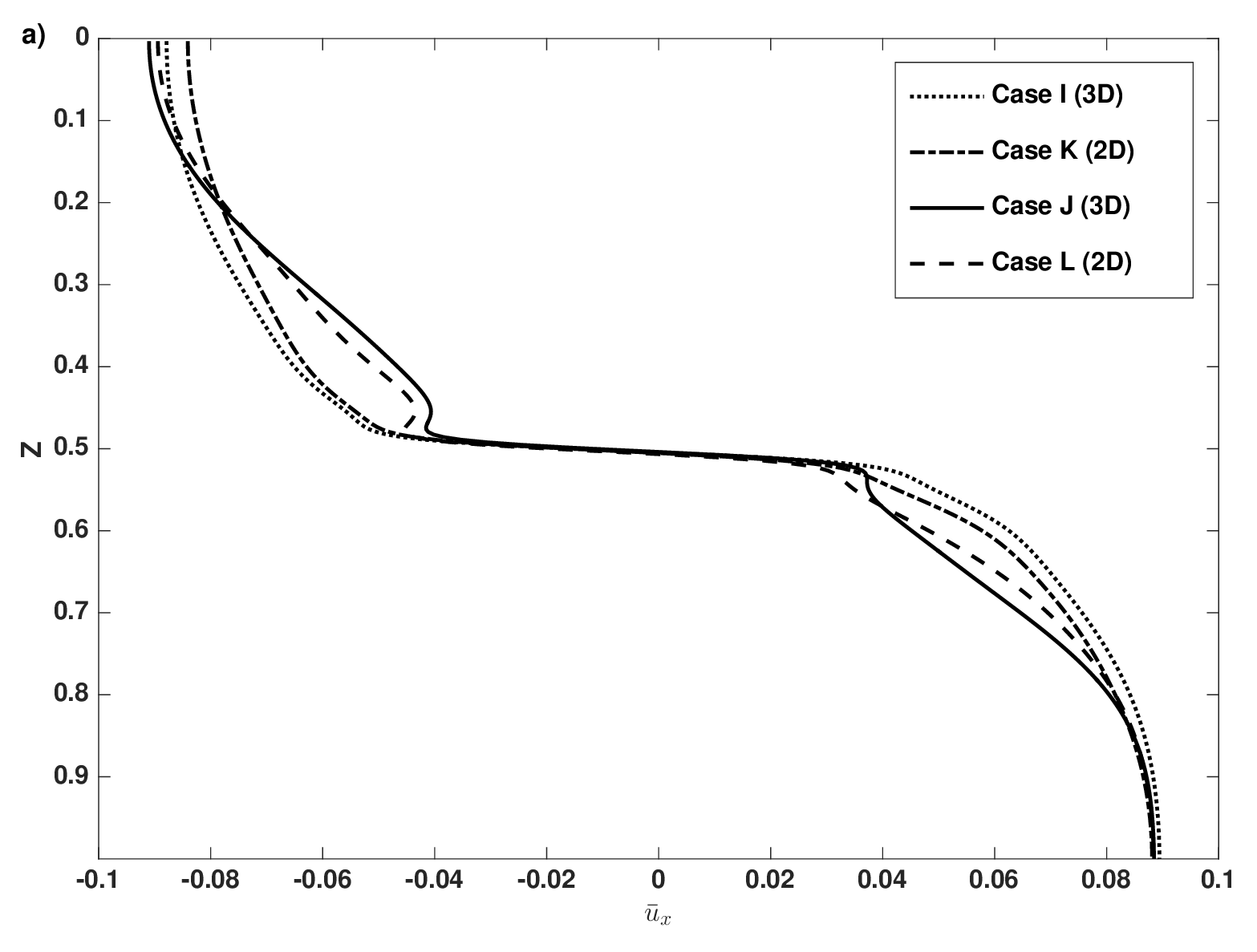}
\includegraphics[width=0.48\textwidth]{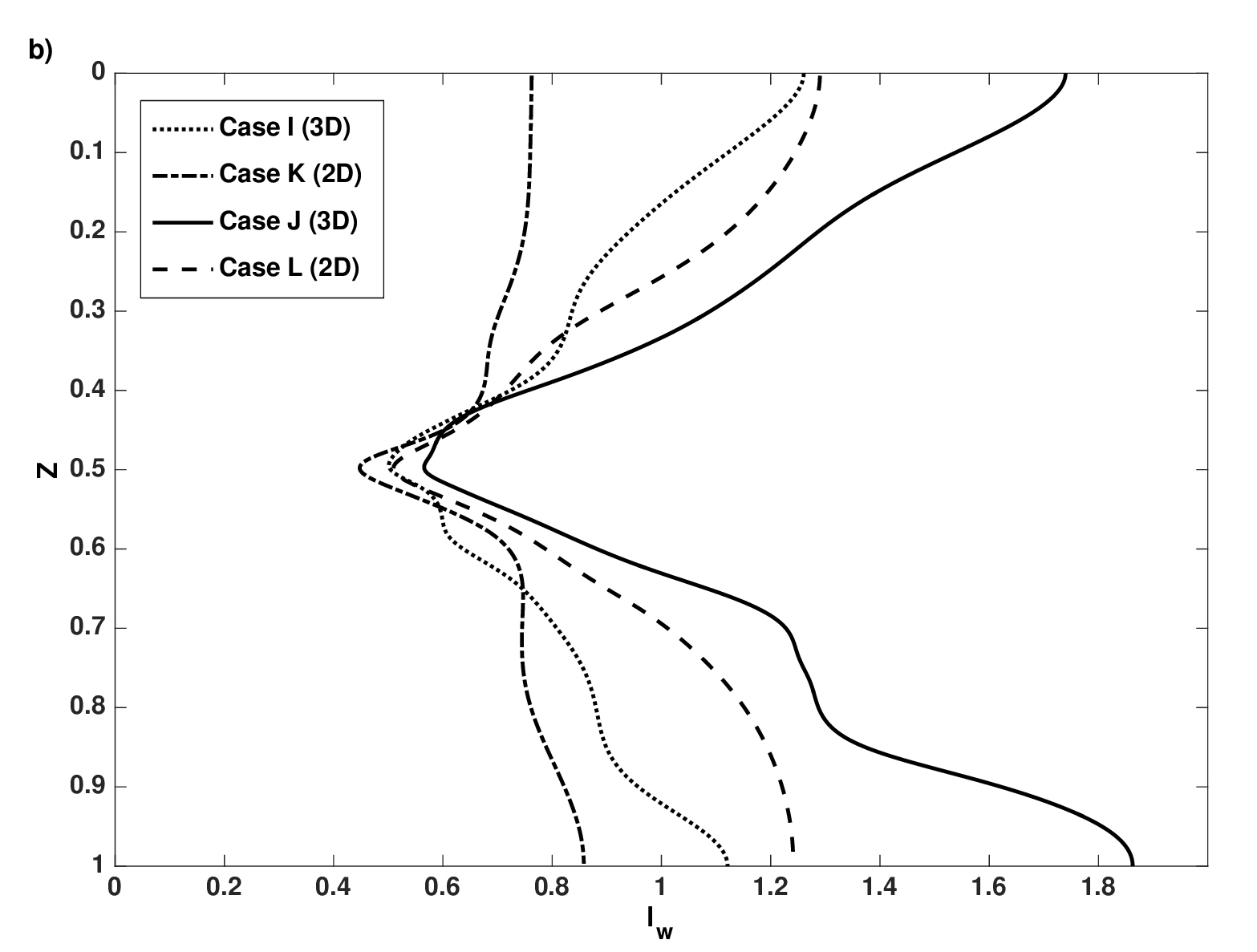}
\caption{ Turbulence characteristics. (a) Horizontally averaged and time averaged $\bar{u}_x$ profiles. (b) Typical turbulent length-scale $l_w$.   }   
\label{fig:figure_3dim_additional} 
\end{figure}
In Fig.~\ref{fig:figure_3dim_additional}~(a) the horizontally averaged velocity in the x-direction is shown for all four cases. While for the low $Pe$ number cases J and L a `staircase like' profile is found, the velocity profile in the large $Pe$ regime is best described as a hyperbolic tangent profile. There is no qualitative difference in the profiles for the two-and three-dimensional calculations. We now compare  the change in the typical length-scale, $\bar{l}_{w}$ (see Table~\ref{table:two_dim_surv}), from the large $Pe$ number case to the low $Pe$ number case obtained in three-dimensional calculations with the corresponding change in two-dimensional calculations. The typical length-scale increases in both two-and three-dimensional calculations. In Fig.~\ref{fig:figure_3dim_additional}~(b) the typical length-scale, $l_w$, is plotted with depth to illustrate the change from the middle of the domain to the boundaries.  However, the values of all quantities obtained during the saturated phase, e.g. the root-mean-square velocity and typical length-scales, are different for the three-dimensional calculations compared to the corresponding two-dimensional calculations. While there are significant differences between all cases, the form of the profile around the middle of the domain is similar. A similar shift towards lower values in the large $Pe$ regime is found for both two-and three-dimensional calculations. The differences towards the upper and lower boundaries for the two-and three-dimensional calculations indicate that there are different dynamics, which is expected due to the existence of secondary instabilities leading to a solution varying in the $y$-direction in three-dimensional calculations. \\
Finally, comparing the changes in the turbulent $Re_t$,  and $u_{rms}$ from the low $Pe$ regime to the large $Pe$ regime in the two-dimensional calculations with the changes obtained in the three-dimensional calculations, summarised in Table \ref{table:two_dim_surv}, we find that the trends are the same. Therefore, we conclude that while there are inevitable differences in the detailed turbulent characteristics obtained in three-dimensional calculations, the overall effect obtained by varying the $Pe$ number is qualitatively the same as in two dimensions. 

%
%
%
\section{Conclusions}
In order to obtain a comprehensive understanding of stars in their entirety, it is crucial to understand the complex dynamics in stellar interiors by investigating shear driven turbulence. Shear driven turbulence is a promising candidate in order to explain the missing mixing problem and is important for magnetic field generation.\\
Numerical calculations are used to obtain a comprehensive insight to the detailed small scale dynamics of shear regions. However, due to computational limitations, all calculations to date use modelling parameters that are far from the actual  values in stellar interiors. Therefore, it is important that we consider how varying key properties affects our understanding of complex stellar regions.\\
In our study we focused on understanding the effect of different viscosities and thermal diffusivities on the saturated phase of a shear driven turbulent flow in a fully compressible polytropic atmosphere. 
Examining the global properties of the saturated flow of an unstable system revealed that the vertical extent of the mixing region  is primarily controlled by the  Richardson number, but the P\'eclet number also plays a key role through the time-scale on which thermal diffusivity acts on the system.  For greater Richardson numbers we find that the vertical spread of the mixing decreases, which also occurs as  $Pe$ is  decreased in the large P\'eclet number limit. In the  small P\'eclet number limit, i.e.~for $Pe < 1$, an opposite trend is observed, where the vertical spread increases with decreasing $Pe$. This increase is due to the high thermal diffusivity that weakens the effectively stratification as soon as the  $Pe$ is less than unity. This weakening effect occurs only in the small P\'eclet number limit. We find that viscosity does not play an important role in the formation of the global shape of the mean flow during the saturated regime. \\
Turbulent flows can be characterised by the typical length-scale and the root-mean-square velocity of the perturbations.  We showed that the typical turbulent length-scales depend on the Richardson number as well as on the P\'eclet number regime.
Investigating strongly stratified systems, and systems in the large P\'eclet number regime, we find that there is a different behaviour in the turbulent characteristics depending on whether it is the Richardson number or  the P\'eclet number that is increased.  While in both cases the product $RiPe$ was increased by the same amount, the system responds differently. 
Therefore, we conclude that the product of the input P\'eclet number and the Richardson number, $RiPe$, can not be used in such strongly stratified systems to extract  information on the characteristics of the turbulence, and both dimensionless number should be provided.
In summation, the turbulent regime of a shear flow instability depends on several parameters and these can counteract each other. The properties of the saturated regime can only be broadly predicted from the input parameters. \\
In the latter part of our research we focused on the low P\'eclet number regime. While for large P\'eclet numbers the initial flow requires low $Ri$ numbers to become unstable, for low P\'eclet numbers it is possible to destabilize a high Richardson number shear flow \citep{1999AA...349.1027L, 2015AandAWitzke}. We examined cases where there were unstable secular shear instabilities. We found in these cases that a different dependency exists, where the typical length-scale increases with decreasing viscosity, which is not the case in large P\'eclet number regimes. \\
Having established a better understanding of what parameters significantly affect the global properties of saturated shear flow instabilities, future studies of the mixing behaviour and momentum transport of shear-driven turbulence can now be conducted. In order to gain a more comprehensive picture of the complex dynamics in stellar interiors, it is crucial to include magnetic field interactions. Future investigations with magnetic fields will help to inform how magnetic fields affect the turbulent regime. Moreover, the results obtained can be used to seek for a shear induced turbulence that is capable to drive a magnetic dynamo, which is subject to current investigations.

\section*{Acknowledgements}
This research has received funding from STFC and from the
School of Mathematics, Computer Science and Engineering at City, University of
London. This work used the DiRAC Data Analytic system at the University of Cambridge, operated by the University of Cambridge High Performance Computing Service on behalf of the STFC DiRAC HPC Facility (www.dirac.ac.uk). This equipment was funded by BIS National E-infrastructure capital grant (ST/K001590/1), STFC capital grants ST/H008861/1 and ST/H00887X/1, and STFC DiRAC Operations grant ST/K00333X/1. DiRAC is part of the National E-Infrastructure. 
\bibliographystyle{mn2e}
\bibliography{bibfile}

\appendix

\label{lastpage}
\end{document}